\documentclass[iop]{emulateapj}
\usepackage{color}

\usepackage{epstopdf}
\usepackage{amsmath}

\makeatletter
\newcommand\lsim{\mathrel{\rlap{\lower4pt\hbox{\hskip1pt$\sim$}}
\raise1pt\hbox{$<$}}}
\newcommand\gsim{\mathrel{\rlap{\lower4pt\hbox{\hskip1pt$\sim$}}
\raise1pt\hbox{$>$}}} 
 
 \newcommand{\tot}{\mathrm{tot}}

 \newcommand{\inner}{\mathrm {in}}
 \newcommand{\out}{\mathrm {out}}

\makeatother
\shorttitle{ Triple Stars'   Mergers and Obliquities }
\shortauthors{Naoz \& Fabrycky}
\makeatother
\begin{document}
\title{ Mergers and Obliquities in Stellar Triples}
\author{Smadar Naoz\altaffilmark{$\dagger$},} 
\affil{Harvard Smithsonian Center for Astrophysics, Institute for
  Theory and Computation, 60 Garden St., Cambridge, MA 02138}
  \altaffiltext{$\dagger$}{Einstein Fellow}
\email{snaoz@cfa.harvard.edu}
\author{Daniel C.~Fabrycky}
\affil{Department of Astronomy and Astrophysics, University of Chicago, 5640 South Ellis Avenue, Chicago, IL 60637, USA}

\begin{abstract}
Many close stellar binaries  are accompanied by a far--away star.
  The ``eccentric Kozai-Lidov" (EKL) mechanism can cause dramatic inclination and eccentricity fluctuations, resulting in tidal tightening of inner binaries of triple stars.
We run a large set of Monte--Carlo simulations including the secular evolution of the orbits, general relativistic precession and tides, and we determine the semimajor axis, eccentricity, inclination and spin-orbit angle distributions of the final configurations.   We find that the efficiency of forming tight binaries ($\lsim 16$~d)  when taking the EKL mechanism into account is $\sim 21\%$, and about $4\%$ of all simulated systems ended up in a merger event.  These merger events can lead to  the formation of blue-stragglers. 
Furthermore, we find that the spin-orbit angle distribution of the inner binaries carries a signature of the initial setup of the system, thus  observations can be used to disentangle close binaries' birth configuration. 
The resulting  inner and outer final orbits' period distributions, and their estimated fraction, suggests secular dynamics may be a significant channel for the formation of close binaries in triples and even blue stragglers. 
\end{abstract}

\section{Introduction}\label{intro}

Most massive stars reside in a binary configuration \citep[$\gsim 70\%$ of all OBA spectral type stars, see][]{Raghavan+10}.  Furthermore, stellar binaries are responsible for diverse astrophysical phenomena, from X-ray binaries to type Ia supernova.   However, probably many of these binaries are in fact  triples.  \citet{T97} showed that
$40\%$ of short period binary stars  ($<10$~d) in which the primary is a
dwarf ($0.5 -1.5\,M_{\odot}$) have at least one additional
companion.  This number contrasts with his estimate of that the fraction of companions to binaries with a slightly larger period ($10-100\,$d), which is $\sim10\%$.  Moreover,
\citet{Pri+06} have surveyed a sample of contact binaries, and noted
that among 151 contact binaries (brighter than 10 mag.), 42$\pm5\%$ are
at least triple. Therefore, triple star systems are probably very common \citep[e.g.,][]{T97,Tokovinin+06,Eggleton+07,Griffin12}. 

Recently, \citet{Rappaport+13} estimated that the fraction of tertiaries within several AU of the close \emph{Kepler} eclipsing binaries is about $20\%$ \citep[see also][]{Conroy+13}. This is in agreement with previous  estimates of triples fraction.  \citet{Rappaport+13}  also reported distribution of mutual inclination angle of their 39 triple-star candidates, which showed a significant  peak around $40^\circ$,  the nominal Kozai angle (see below).

 From dynamical stability arguments 
these triple stars must be in hierarchical configurations, in which the inner binary is
orbited by a third body on a much wider orbit.  
Many short--period compact binaries \citep[such as Black Holes, Neutron stars and White Dwarfs; ][]{Tho10}  are likely
produced through triple evolution.  Secular effects (i.e., coherent
interactions on timescales long compared to the orbital period), and
specifically Kozai-Lidov cycling \citep[][see below]{Kozai,Lidov},
have been proposed as a dynamical driver in the evolution of triple
stars \citep[e.g.][]{Har69,Mazeh+79,1998KEM,Dan,PF09,Tho10,Naoz+11sec,Shappee+13,Pejcha+13,Perets14,Michaely+14}.  In addition,
Kozai-Lidov cycling speed the growth of black holes at the centers of dense star clusters
and the formation of short-period binary black holes
\citep{Wen,MH02,Bla+02}.  In addition, \citet{Iva+10} estimated that the
most efficient formation channel for black hole X-ray binaries in globular
clusters may be triple--induced mass transfer in a black hole--white
dwarf binary.

In early studies of high-inclination secular perturbations
\citep{Kozai,Lidov}, the outer orbit  was assumed to be circular and that  one of the inner binary members  is a test (massless) particle. 
 In this situation, the
component of the inner orbit's angular momentum along the z-axis  is
conserved (where the z-axis is parallel to the total angular momentum), and the lowest order of the approximation, the quadrupole approximation, is valid.
 Following \citet{LN} we label this approximation as the ``TPQ" (Test Particle Quadrupole) approximation\footnote{ Note that here we focus on the general problem with no restrictions on the masses, for the test particle approximation see \citet{LN}, \citet{Boaz2} and \citet{Li+14}.}. Recently, \cite{Naoz11,Naoz+11sec} showed that relaxing  these assumptions leads to qualitative different behavior. Considering systems beyond the test particle approximation, or an eccentric orbit with a moderate semi-major axis ratio, requires the next level of approximation, called the octupole--level of approximation \citep[e.g.][]{Har68,Har69,Ford00,Bla+02}.

In the octupole level of approximation, the inner orbital eccentricity can reach very high values \citep{Ford00,Naoz+11sec,Li+13,Tey+13}.  In addition, the inner orbit inclination with respect to the total angular  momentum can flip from prograde to retrograde --- not even the sign of the orbital angular momentum's z-component is conserved \citep{Naoz11,Naoz+11sec}.  We refer to this process as the {\it Eccentric Kozai--Lidov} (EKL) mechanism. It was shown in \citet{Naoz+11sec} that the secular approximation can be used as a great tool to understand different astrophysical settings, from massive or stellar compact objects to planetary systems \citep[for example this has large consequences on retrograde hot Jupiters, e.g.,][]{Naoz11}. 

We study the secular dynamical evolution of triple stars using the octupole level of approximation, including tidal effects \citep[following][]{Hut,1998EKH} and general relativity (GR) effects for both the inner and outer orbit  \citep{Naoz+12GR}\footnote{Note that the 1PN interaction term, between the inner and outer orbit, mentioned in  \citet{Naoz+12GR} has negligible effect here.   }. The secular evolution of triple stellar system was considered previously in the literature  \citep[e.g.][]{Har68,Har69,Mazeh+79,1998KEM,1998EKH,Egg+01,Egg+06,Dan}. Specifically we point out that \citet{Dan} ran large Monte-Carlo simulations for the evolution of triple stars including  GR and   tidal effects using the quadrupole test particle approximation. Here we show that the octupole--level of approximation can result in additional behavior, where we focus on the following items: the formation of short--period binaries, the obliquity of inner binaries, merged systems and the outcome of the outer orbit. We provide comparisons with observed catalog and known systems when possible.  The octupole level of approximation can lead to very high eccentricities that can drive the two members of the inner to collision (or grazing interactions). This was noted first by \citet{Ford00} and later suggested to be important to white dwarf binaries by \citet{Tho10},  and here we quantify this including tidal interactions.

The merger between the two inner members due to the large eccentricities induced by the octupole level has been suggested recently as a possible mechanism to explain double degenerate type Ia supernova \citep[e.g.,][]{Tho10,Hamers+13,Prodan+13}. In this scenario a third body in the system induces  large eccentricity that drives the inner binary to a near radial trajectory.  Moving beyond the secular approximation, triple body dynamics seems to still plays a dominate role in causing the collision of two white dwarfs and result in type Ia supernova \citep[e.g.,][]{Katz+12,Kushnir+13,Dong+14}.
There are at least two points that can affect this outcome. 
  First, the triple population should have undergone EKL evolution before the white dwarf stage. Second, stellar evolution, and especially mass loss can play important role in the evolution of these systems \citep[e.g.][]{PK12,Shappee+13}, as it will tend to expand the orbits, or even produce unbound objects \citep{Veras+12}. 
 Here we present results relating to the first part, where we follow the  secular evolution of triple stars. 
 We find the parts of the parameter space that has already undergone EKL evolution and resulted in either close system or systems that crossed their  Roche limit. These systems have decoupled  from   the third object, and thus will probably not be a part of the parameter space that is available for the double degenerate scenario.

The formation of blue stragglers has also been discussed in connection to dynamics in triples.  \citep{PF09} suggested that the close binaries created by Kozai cycles with tidal friction would then merge by losing their orbital angular momentum to magnetized winds.  \citep{aaron+11,LG13} on the other hand studied encounters of multiple-body systems in star clusters, and found direct collisions to be a possible source of blue stragglers.   Our work emphasizes the efficiency of collisions when the octupole-level interaction is taken into account, such that blue stragglers can result from prompt collisions, even for isolated triples. 

Another aspect of the octupole level is that it results in a qualitative different  time evolution of the obliquity, the angle between the star spin axis and the binary orbit \citep{Naoz11,Naoz+12bin}. As more binary stars obliquities are being observed for example by the BANANA survey \citep{Banana}, and by other individual endeavors, we give specific predictions for the obliquities of inner binaries in triples.  Below we compare our results to the current available observations, and find them to be consistent with our secular model. 

The paper is organized as follow, we begin by describing the numerical setup (\S \ref{ICs}). We present our results (\S \ref{results}) and focus on  the effects of tidal dissipation and binary merger (\S \ref{sec:close_inner}), the mutual inclination and obliquity (\S \ref{sec:Inc_Spin}) and specific analysis of the outer orbit configuration (\S \ref{sec:outer}). We discuss our results and predictions in \S \ref{sec:dis}.

\section{Numerical Setup}\label{ICs}

We follow the numerical setup presented in \citet{Dan}. 
We set $m_1=1\,M_\odot$ $m_2$ was chosen by selecting the $q_{\inner}=m_2/m_1$ from a  Gaussian distribution with mean of $0.23$ and standard deviation of $0.42$  \citep{Duquennoy+91}. Similarly,  $m_3$ was set by choosing  $q_{\out}=m_3/(m_1+m_2)$ from the same  Gaussian distribution.  This way enables calibration to different choices of initial mass function.  As we will show, the final results  only weakly depend on the mass ratio, and thus we expect that choosing different mass ratio distribution will have little effects. 
We denote the inclination angle of the inner (outer)
orbit with respect to the total angular momentum by $i_1$ ($i_2$), so
that the mutual inclination between the two orbits is
$i_\tot=i_1+i_2$.  
We draw the inner (outer) periods, $P_{in}$ ($P_{out}$), from the log-normal distribution of
\citet{Duquennoy+91}.    Note that  this period distribution represent the final periods of binaries population, rather then initial one. Furthermore, there is not a clear evidence that this distribution is the correct initial distribution for triples. We choose this distribution for self consistency reasons, and as we will show, even in light of these  caveats, comparing our results to observations and catalogs suggests  that the EKL mechanism plays an important role in triples.   
 The  distribution of the inner and outer
eccentricities ($e_1$ and $e_2$ respectively) was chosen to  uniform, following   \citet{Raghavan+10}.  This distribution is more conservative than a thermal distribution, i.e., uniform distribution yields less eccentric outer binaries and since the EKL mechanism is more efficient for larger eccentricity we are considering a more conservative case.

We then require these initial conditions satisfy dynamical stability, such that we can separate the effect of long-term secular effects.  The first condition is that the inner orbit is initially outside the Roche limit, lest the inner stars suffer a merger before the tertiary can act.  The second condition is long-term stability of the triple, in which we follow the \citet{Mardling+01} criterion:
\begin{equation}\label{eq:Mar} \frac{a_2}{a_1} > 2.8\left(1+\frac{m_3}{m_1+m_2}\right)^{2/5}\frac{ (1+e_2)^{2/5}}{ (1-e_2)^{6/5}} \left(1-\frac{0.3 i_{\tot}}{  180^\circ}\right)      \end{equation}
A final criterion is:
\begin{equation}\label{eq:epsilon}
\epsilon=\frac{a_1}{a_2 }\frac{e_2 }{1-e_2^2}<0.1 \ ,
\end{equation}
where $\epsilon$ measures the relative amplitudes of the octupole and quadrupole terms in the triple's Hamiltonian.  This is numerically similar to the stability criterion, Equation~(\ref{eq:Mar}) \citep[as shown in][]{Naoz+12GR}.  At the extreme of inequality~(\ref{eq:epsilon}), effects beyond the octupole may dominate the dynamics  -- \citet{Katz+12},  \citet{Antognini+13},  \citet{Antonini+14} and \citet{Bode+14}  have shown that the secular approximation fails for a strong perturber -- however, the equations we adopt do not describe these situations.  Our justification for ignoring these effects is that inequalities (\ref{eq:Mar}) and (\ref{eq:epsilon}) are very similar numerically, meaning that there are few systems which are stable yet are poorly described by the octupole approximation. 

We solve the octupole--level secular equations numerically following
\citet{Naoz+11sec}. We also include General Relativity precession for the inner and outer orbit following \citet{Naoz+12GR}. We are able to follow the spin vectors of both the primary and the secondary of the inner orbit stellar components 
 (spins of both stars are set initially  to $25$~d).  
Specifically we are interested in the angle between the angular momentum of the inner orbit and the spin of the stars (the spin--orbit angle, $\psi$), which was set initial on a uniform distribution for the primary while  the secondary was set initially with $\psi=0^\circ$ (but we also investigated other configurations, see  \S \ref{sec:Inc_Spin}).

We also include tidal interactions for the inner binary evolution. 
The differential equations that govern the inner binary's tidal
evolution were presented in \citet{Egg+01}. These equations take into
account stellar distortion due to tides and rotation, with tidal
dissipation based on the theory of \citet{1998EKH}.  The viscous time
scale, $t_{V}$ (set to be $5$~yr in all of our runs), is related to the quality factor $Q$ \citep{GS66,Hansen10} by
\begin{equation}
Q_{j}=\frac{4}{3} \frac{k_{L,j}}{(1+2 k_{L,j})^2}\frac{G m}{R_{j}^3}\frac{P_{in}}{2 \pi} t_{V,j} \ ,
\end{equation}
where $j\in \{1, 2\}$ for $m_1$ and $m_2$ respectively, and $k_{L}$ is the
classical apsidal motion constant.  We use the typical value $k_{L} =
0.014$, valid for $n = 3$ polytropes, when representing stars  \citep{Egg+01,Dan}.  Note that this approach simplifies the effects of tides, since one would expect that that the viscous time scale would vary with the stellar mass and radius, however, the exact dependence is unknown. 

 The strength of the equilibrium tide recipe used here  is that  it is self consistent  with the secular approach used throughout this study. Furthermore, assuming polytropic stars this recipe has only one dissipation parameter for each star. Using this description we are able to follow the precession of the spin of the stars  due to oblateness and tidal torques. Therefore, using this approximation enables a qualitative understanding of the physical effects in the system.  When the stars are in a very close  pericenter passage (eccentricity approaches unity), the equilibrium tides model breaks as higher orders modes in the stars may play larger roles, which can affect the dynamical evolution, however, this is beyond the scope of this paper.

The upper limit for each system's integration time in all our
simulations was $10\,$Gyr.  When the two inner stars are tidally captured the integration becomes extremely
expensive. Therefore,  we adopt a stopping condition which satisfy that both $e_1\leq5\times 10^{-5}$ and $P_{in}\leq 7$~d.

The EKL mechanism can cause very large eccentricity excitations for the inner orbit,  implying a high probability that the stars  will cross each other  the Roche
limit.  Following \citet{Eggleton83}, we define the dimensionless    number: 
\begin{equation}\label{eq:Roche}
L_{Roche,ij}=0.49\frac{ ({m_i} /{m_j})^{2/3} }{0.6 (m_i/m_j)^{2/3}+\ln (1+(m_i/m_j)^{1/3})} \ ,
\end{equation}
where $i,j\in \{1, 2\}$.   Note that this criterion is for circular orbits and does not necessarily  describe the full dynamics of the system. Furthermore, this condition is only slightly (up to a factor of unity) less conservative than the simplified relation of $\sim 0.3 (m_i/m_j)^{1/3} $. However, we use this as a qualitative criterion for our purposes.

In many of our simulations,
the inner orbit reaches extremely high eccentricities. 
 During
excursions to high eccentricity there is a competition between the
increased efficiency of tides leading to the Kozai capture process
\citep{Naoz11} and the possibility of destroying the system by
crossing the Roche limit (see below for further discussion). To address the possibility of crossing the Roche limit we set an additional stopping condition, which satisfy that if $a_1 (1-e_1) L_{Roche,ij}< R_i$ we stop the
run and assume that the inner binary  merged.

\begin{table}
 \caption{Summary of the simulations}
\label{table_sim}\vspace{-0.7cm}
\begin{center}
\begin{tabular}{l  c | c c | c c }
\hline
name & $\#$ of  &   $\psi_{1,IC}$ &  $\psi_{2,IC}$ & close  &   Roche \\
    &    systems       &                          &                         &      binaries                 & limit\\
    \hline
    EKL & 3050 &   uniform & $\psi_{2,IC}=0^\circ$ &   $21\%$& $4\%$ \\ 
     EKL$\psi90$ & 1141 &   $\psi_{1,IC}=90^\circ$  & $\psi_{2,IC}=90$ &   $21\%$& $4\%$ \\ 
       EKL$\psi0$ & 1139 &   $\psi_{1,IC}=0^\circ$  & $\psi_{2,IC}=0$ &   $21\%$& $4\%$ \\ 
TPQ & 2103 &   uniform & $\psi_{2,IC}=0^\circ$ & $16\%$ & $2\%$ \\
\end{tabular}
\end{center}
\end{table}

  We also compare our results with \citet{Dan} by running an additional set of Monte-Carlo simulation while considering only the TPQ approximation. The only major difference in our setting of the TPQ runs compared to \citet{Dan} is that the initial eccentricity distribution is uniform \citep{Raghavan+10}. Furthermore, in our initial period distribution we only allow system that are above the Roche limit separation. The total number of systems we have run are specified in table \ref{table_sim}.

\begin{figure*}[!t]
\begin{center} 
\includegraphics[width=0.8\linewidth]{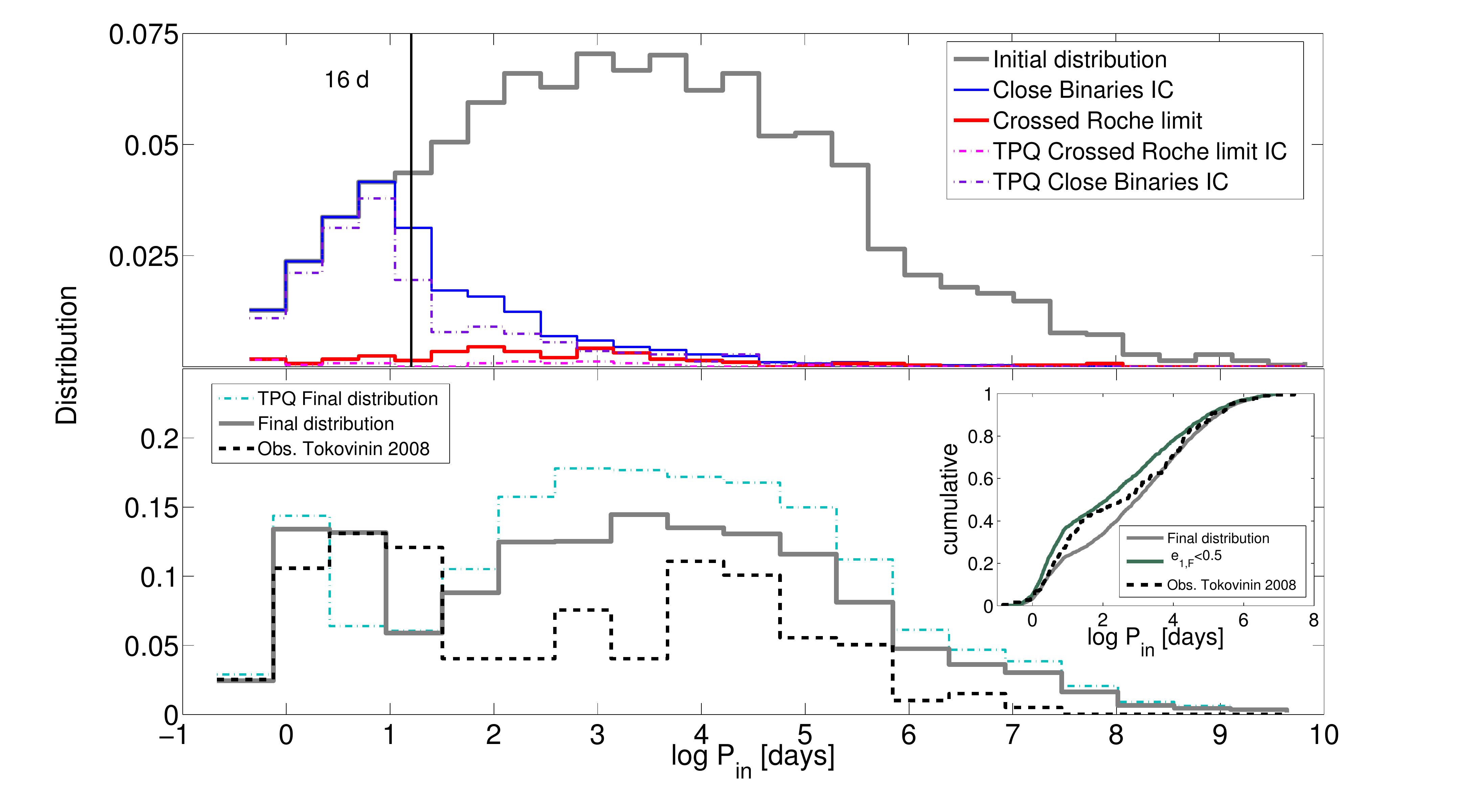}
\caption{The initial ({\bf top panel}) and final ({\bf bottom panel}) distribution of the inner orbit period. {\it In the top panel} we consider the initial distribution of  all of the runs (grey line), the binaries that ended up in close configuration (blue line) and those systems that crossed the Roche limit during their evolution (red line); see text for more details.  We also consider the initial distribution {\it for the TPQ case} of the binaries that ended up in close configuration (purple dash-dot line, slightly off set for visualization purposes) and crossed the Roche limit (pink  dash-dot line). The distributions are scaled such that the integral of the initial distribution is unity. {\it In the bottom panel} we show the final distribution of all of the runs for the EKL mechanism  (grey line) and the TPQ case (cyan dash-dot line).  We also show,  the observational distribution of  inner orbit systems in triples taken from \citet{Tok08}, for systems closer than $50$~pc (black dashed line). The theoretical distributions are  scaled to  the observed distribution. {\it In the inset} we show the cumulative distribution of the observations distribution  taken from \citet{Tok08} (black dashed line), compared to the final distribution (grey solid line). Furthermore, since the observations have typical inner orbital eccentricity of $0.5$, we show the final distribution of system with $e_{1,F}<0.5$ (green solid line).  
 \label{fig:Pdis} }\end{center}
\end{figure*}

\section{Results}\label{results}

\subsection{The Inner Binary Period Distribution}

Similarly to \citet{Dan} we find that the final inner orbital period's distribution is consistent with a bimodal distribution with a peak  around $\sim3$~days, as shown in Figure  \ref{fig:Pdis}. We define ``close binaries'' as systems with final period shorter than $16$~d; this period roughly separates the two mayor peaks. We also show in the bottom panel of this Figure, the   observed period distribution of  inner binaries  in triples adopted  from \citet{Tok08} public catalog (where we have scaled the theoretical  distribution to  the catalog's one to guide the eye).  The inset in the bottom panel shows the cumulative distribution of the simulated runs as well as the observed data. The observed systems in the catalog  have typical inner binary eccentricity of about $0.5$,   \citep{Tok08}. Taking this at face value, which may point out to some selection effect in the catalog, we have compared the observed cumulative distribution to our calculated distribution, limiting our final inner orbital eccentricity, $e_{1,F}<0.5$ (green solid line). This yields better agreement between the simulated period distribution and the one taken from the catalog.  We note that near the completion of this paper \citet{Tokovinin14b,Tokovinin14a} reported a new database of triple stars, here we use his old data base, which mainly differ in sample size.

Interestingly, the observed bimodal distribution is reproduced with our secular evolution model.  We find that  the  Kolmogorov-Smirnov test does not reject the null hypothesis 
that the observed inner orbit's distribution and the simulated one are from the same continuous distribution (where we consider the full EKL distribution, and not the eccentricity limited one), with $p$ value of $0.3828$. This behavior  suggests that secular evolution in triple's plays an important role in shaping the distribution of these systems.  
However, we point out two major differences between the theoretical predictions and the observational data. 

The first is the period which separates between the two major peaks in the distribution. The  observed period distribution (adopted  from \citet{Tok08} public catalog) have a wide valley in the distribution  with periods ranging  between  $25-100$~d while the theoretical predictions gives a narrow valley in the period distribution ranging between  $16-40$~d.  The explanation may lay either on the initial conditions, where we assumed that hierarchical triple period distribution follows binary population,   or in our model.     Here we restricted ourselves to  the hierarchal three-body approximation which means that systems that could have formed through  short timescales, scattering-type of  interactions are not modeled. 
 Wide inner orbits in  triple configurations have been found in  scattering-like interactions between stars in an open stellar cluster in the recent work by  \citet{Geller+13}, where we deduce, from their figure 9,  a minimum in the period distribution that extended to $\sim 100$~d (note that their second, long period peak in the distribution is not as  dominant as in our case for the triple population). An additional process that can produce wide inner binaries is mass loss \citep{PK12,Geller+13}. With our model we do not capture these possible process that can account for wider inner binaries, therefore, even in the absence of these processes, the agreement between the observed and modeled distributions suggest that secular evolution plays a dominant  role for triple systems.  
 
 The above difference in the period distribution valley between the observation and the TPQ calculation in \citet{Dan} was noted  by \citet{Tok08}. Our new calculation including the octupole evolution improves on that and extends the inner period distribute peak to $\sim 16$~d, where we addressed possible causes for the differences above. However, we are encouraged that the theoretical inner orbital period distribution appear rather flat, as   found in  \citet{TS02} radial velocities  observations.  
 
 The second discrepancy between the observations and the theoretical predictions is in the ratio between the two peaks; in other words at face value, there are more close inner binaries than wide ones in the catalog of the  observed systems compared to the simulated triples. This can be explained as a combination of both theoretical and observational shortcomings. Here again, stellar evolution may play a role in determining the final separation of  triple configurations, and may even ionize/unbind the outer orbit, as shown in \citet{PK12} and \citet{Geller+13}. As shown in Figure  \ref{fig:Pdis} top panel, there is a large tail of wide inner binaries that can result in close binaries. Due to the hierarchical criterion, their tertiary has even a wider orbit. The capture into a close binary due to the EKL mechanism can happen on a short time scale  (as can be seen in Figure \ref{fig:outer} top left panel), which leaves enough time for fly-by perturbation or mass loss to unbind the outer orbit.  This is also supported by \citet{Tokovinin+06} observations of tight binaries without a tertiary.   This may imply  that, in our model, we  over estimate the number of wide binaries (since they are more likely to unbind). 
 
 On the other hand, wide inner binaries are harder to observe and thus the catalog of the  observed systems is incomplete   and may  suffer from some selection effects (A. Tokovinin private communication).   The latter may be supported by the comparison with the $e_{1,F}<0.5$ final distribution (inset of Figure \ref{fig:Pdis}).  These systems better agree with the observed distribution which has    typical inner eccentricity of $0.5$  as noted in \citet{Tok08} and \citet{TS02}. This suggests that the either we are missing a piece of the physics or that observations are biased against high eccentric systems. Although this conclusion does not provide a definite answer it points toward a possible explanation.  

The   similarity of the  period's distribution with \citet{Dan} is not surprising since the bulk behavior  did not change by solving the equation of motions up to the octupole  level of approximation.  This means that both the TPQ and the EKL mechanisms produce double peak final period distribution, see Figure \ref{fig:Pdis}. However, in  the presence of the octupole level, larger parts of the phase space are accessible for large inclination and eccentricity oscillations. Most importantly, the inner orbit's eccentricity may reach much larger values than the values reached with the quadrupole approximation.  Therefore, wider inner binaries (compared to the TPQ approximation) and even  lower inclination systems can end up in forming close binaries or even drive the members of the inner orbit to merge (see Figure \ref{fig:Pdis} for specific comparison), which affect the fraction of final close and merged systems  (see Table \ref{table_sim}). 

  Another difference between the EKL and the TPQ approximations is  the location of the minimum in the period distribution.   For the TPQ this is located between $\sim 3-31$~d, while the EKL mechanism yields wider periods with a range between $\sim 8 - 31$~d. This is because in the EKL mechanism, the maximum eccentricity value can vary  rapidly, and can reach extremely  large values. A system that reached a large eccentricity value can result in shrinking  $a_1$ to some other, lower value, in a step like way \citep[see for example  Figure 2 in][]{Naoz11}. This new value,  associated with  lower $\epsilon$ value, may be less favorable for another eccentricity spike, resulting in a stable configuration on a wide inner orbit. On the other hand, the TPQ mechanism produces always the same eccentricity value, which, if high enough it can cause $a_1$ to shrink in a smooth way \citep[see for example Figure 1 in][]{Dan}.   

\begin{figure}[t!]
\includegraphics[width=\linewidth]{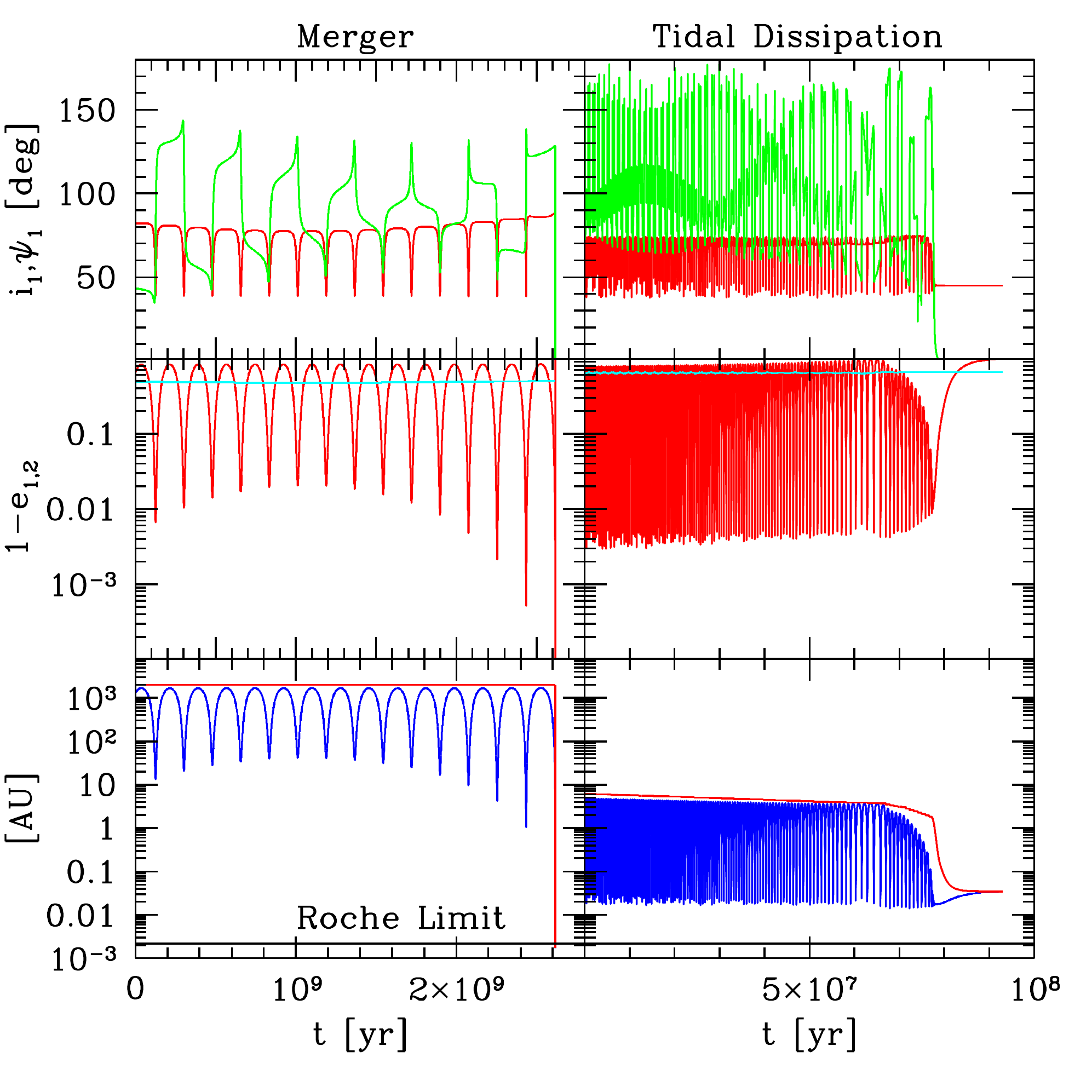}
\caption{We show two examples chosen from the Monte-Carlo runs. The left column shows a system that resulted in merger while the right column is of an inner binary that shirnked its separation  to a stable tight configuration. In the top panel we show the inclination of the inner binary with respect to the total angular momentum $i_1$ (red lines), and the spin orbit angle of the primary (green angle). Middle panel shows the eccentricity as $1-e$ for the inner binary (red lines) and the outer binary (cyan lines). Bottom panel shown the inner binary separation  (red lines) and the pericenter distance (blue lines). We also show the Roche limit value (black lines)  which is simply $L_{Roche,12}\times R_1$.  Left column was set initially with $m_2=0.337$~M$_\odot$, $m_3=0.094$~M$_\odot$, $a_1=6.16$~AU, $a_2=106.155$~AU, $e_1=0.539$, $e_2=0.368$, $\omega_1=223.54^\circ$, $\omega_2=212.863^\circ$ and $i=103.02^\circ$, which means that $i_1=68.52^\circ$ and the  system in the right column was set initially with $m_2=0.31$~M$_\odot$, $m_3=0.733$~M$_\odot$, $a_1=2001.67$~AU, $a_2=31571.32$~AU, $e_1=0.356$, $e_2=0.51$, $\omega_1=145.99^\circ$, $\omega_2=65.82^\circ$ and $i=88.41^\circ$, which means that $i_1=82.14^\circ$. 
} \label{fig:Ex} 
\end{figure}

\subsection{Binary Merger and Tidal Dissipation}\label{sec:close_inner}

During the system evolution,  the octupole--level of approximation can cause large  eccentricity excitations for the inner orbit. Thus, on one hand   the nearly radial motion of the binary  drives the inner binary to merge, while on the other hand  the tidal forces tends to shrink the separation and circularize  the orbit.  If during the evolution the tidal precession timescale (or the GR timescale) is similar to that of  the Kozai time scale, further eccentricity excitations are suppressed \citep[this was already noted in ][ for the quadrupole approximation]{Hol+97,Dan}\footnote{ Note that when GR timescale are similar to the quadrupole--level of approximation timescale a resonant like behavior can occur \citep[e..g, ][]{Naoz+12GR}.}. In that case  tides can shrink the binary separation and form a tight binary decoupled from the tertiary companion. The final binary remained a stable orbit (note that tides  always tend to shrink the binary separation, but this happens on much longer timescale).  However,  if the eccentricity is excited  on a much shorter timescale than the typical 
extra   precession timescale (such as tides, rotational bulge, and the GR precession timescales, but of course still long enough so the secular approximation is valid), the orbit becomes almost radial, and the stars can cross the Roche--limit. In that case, the 
extra precession does not have enough time to affect the evolution. 
This is the process that causes, for example, tidal migrations of planets in stellar binaries until they tidally disrupt or merge \citep{Naoz+12bin}. 
In our Monte-Carlo about $4\%$ of the systems crossed their Roche-limit and about $21\%$ are on a tight ($<16$~day) orbit. Hereafter we label all systems with inner orbit configuration with final period binary  $<16$~d as ``close binaries".


\begin{figure}[btb]
\includegraphics[width=\linewidth]{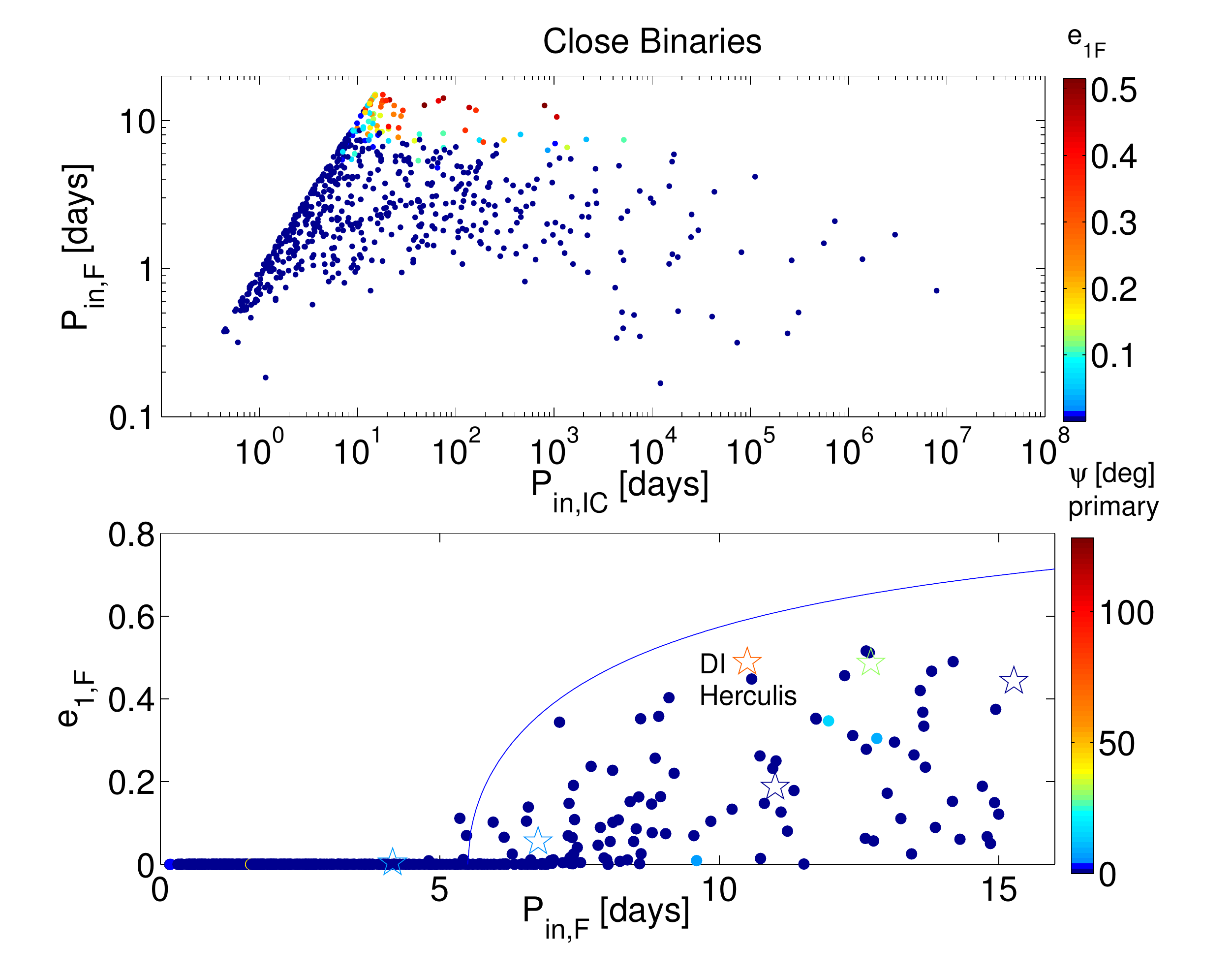}
\caption{
Close Binaries configuration (we consider only the close-in binaries, $P<16$~days). {\bf Top panel:} final inner binary periods as a function of the initial one. The different colors mark the final eccentricity of the system.  {\bf Bottom panel:}  we show the final eccentricity achieved as a function of the final period. The color code is the final spin-orbit angle of the primary, where most of the systems have final spin-orbit lower than $10^\circ$ (see Figure \ref{fig:spinOrbit} for complementary presentation of this parameter space). Note that we show here only the primary's obliquity.  The solid line represent a constant angular momentum curve with a final binary period of $5.5$~days. We also plot the observations (stars symbols) \citep{Albrecht09,Albrecht11,Albrecht13,Triaud+13,Harding+13,Zhou+13}.  Note that the horizontal axis in the bottom panel is in linear scale and for the top panel is in log scale. }
   \label{fig:P1FP0} 
\end{figure}

In Figure \ref{fig:Ex} we show two representative examples of the evolution of two systems that have undergone dramatic changes during their evolution due to close pericenter passages. In general if the eccentricity excitation is evolving gradually (though still to larger values than achieved with the quadrupole--level approximation) the pericenter distance shrinks slowly and  allows tides to work (as shown in the example at the right column).    However, during the evolution shown in  the left column in Figure  \ref{fig:Ex}, the pericenter distance due to the EKL evolution changes by many orders of magnitude from one quadrupole time scale to the next. Thus, the angular momentum of the inner orbit is decreasing by more than an order of magnitude.  The separation shrinks dramatically on short time scale, though still larger than the inner orbit period (more than a factor of 10), and the eccentricity is decreasing on that time scale too.
This happens when the inner orbit reaches large eccentricity while still on a wide separation, in that case the tidal forces  (and the GR) precession  do not have the time to stabilize the system and the binary components  crossed each other Roche-limit.

\begin{figure}[t!]
\includegraphics[width=\linewidth]{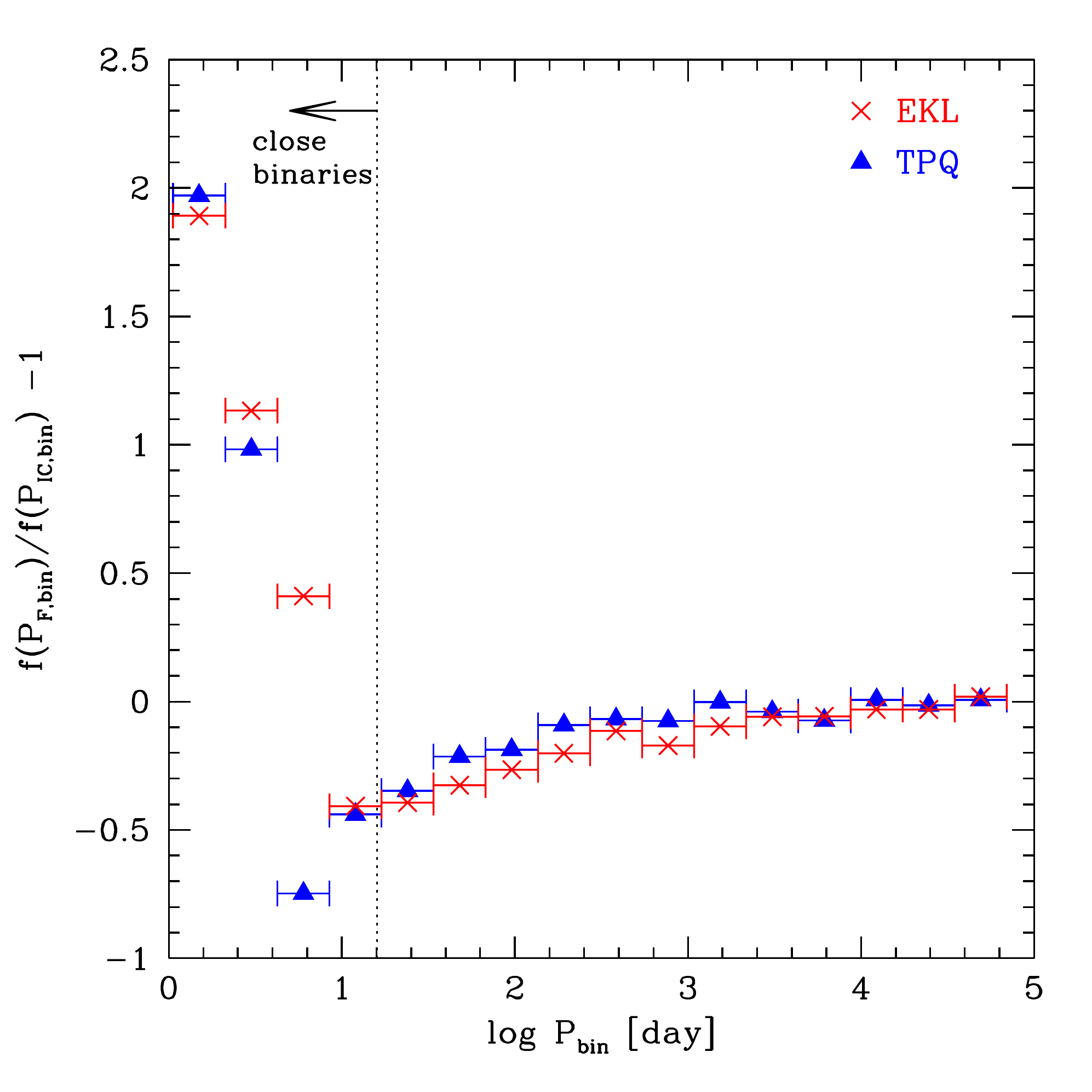}
\caption{The fraction of systems that ended up in an inner period bin
$P_{F,bin}$  period relative to the initial fraction of systems in that
bin. We compare between the EKL (red crosses) and TPQ (blue
triangles).  
} \label{fig:EKLRPQ} 
\end{figure}

In the top panel in Figure \ref{fig:Pdis} we show the initial inner orbit period distribution of the systems that merged during the evolution as well as those that  ended up in close configurations ($<16$~d). As shown in this Figure, the different outcomes (i.e., close binaries or merger) originate from two distinct populations. On average merged binaries (marked in red) are more likely to originate from initially wider inner  binaries (we discuss the outer orbit configurations for these system in \S \ref{sec:outer}).
 
In Figure \ref{fig:P1FP0}, top panel, we consider the relation between initial and final period of the inner close binaries' population ($<16$~d). 
As shown in this Figure (and also in Figure \ref{fig:Pdis}), the main contribution of the close binaries' population (associated with peak in the period distribution of $\sim 3$~d) comes from systems with inner binaries with periods of $\sim 4 - 16$~d. However, about $\sim 41\%$ of the close binaries originated from initial inner binaries separation larger than $16$~d.  Considering the entire triple population, we find that about $8.6\%$ of {\it all} triples with initially $P_{in}>16$~d have become close binaries (i.e., with $P_{in,F}<16$~d). Comparing this number with the TPQ's of $\sim 3.6\%$, we find that the EKL efficiency is larger by about $2.4$ compared to the TPQ.  To illustrate further the difference between the EKL and TPQ we
consider in Figure \ref{fig:EKLRPQ} the fraction  systems in a
final inner period bin relative to the initial systems in that bin (for
equal logarithmic close binaries bins). The period valley in the
distribution is indicative of the different efficiency of the EKL and
TPQ approximations.

The population of close binaries with a wider final separation  have a non-negligible eccentricity and is in the process of tidal shrinking and circularization, as shown in the bottom panel of Figure \ref{fig:P1FP0}.  Interestingly,  this bottom panel  is qualitatively similar to figure 2 in \citet{TS02}. The tidal shrinking process is  still on its way even after  $10$~Gyr of integration time, and those systems lay under the constant angular momentum curve (see also Figure \ref{fig:outer}, top left panel that shows the integration  time). 
In section \ref{sec:outer} we discuss the outer orbit configuration for those close systems.

\begin{figure}[!t]
\includegraphics[width=\linewidth]{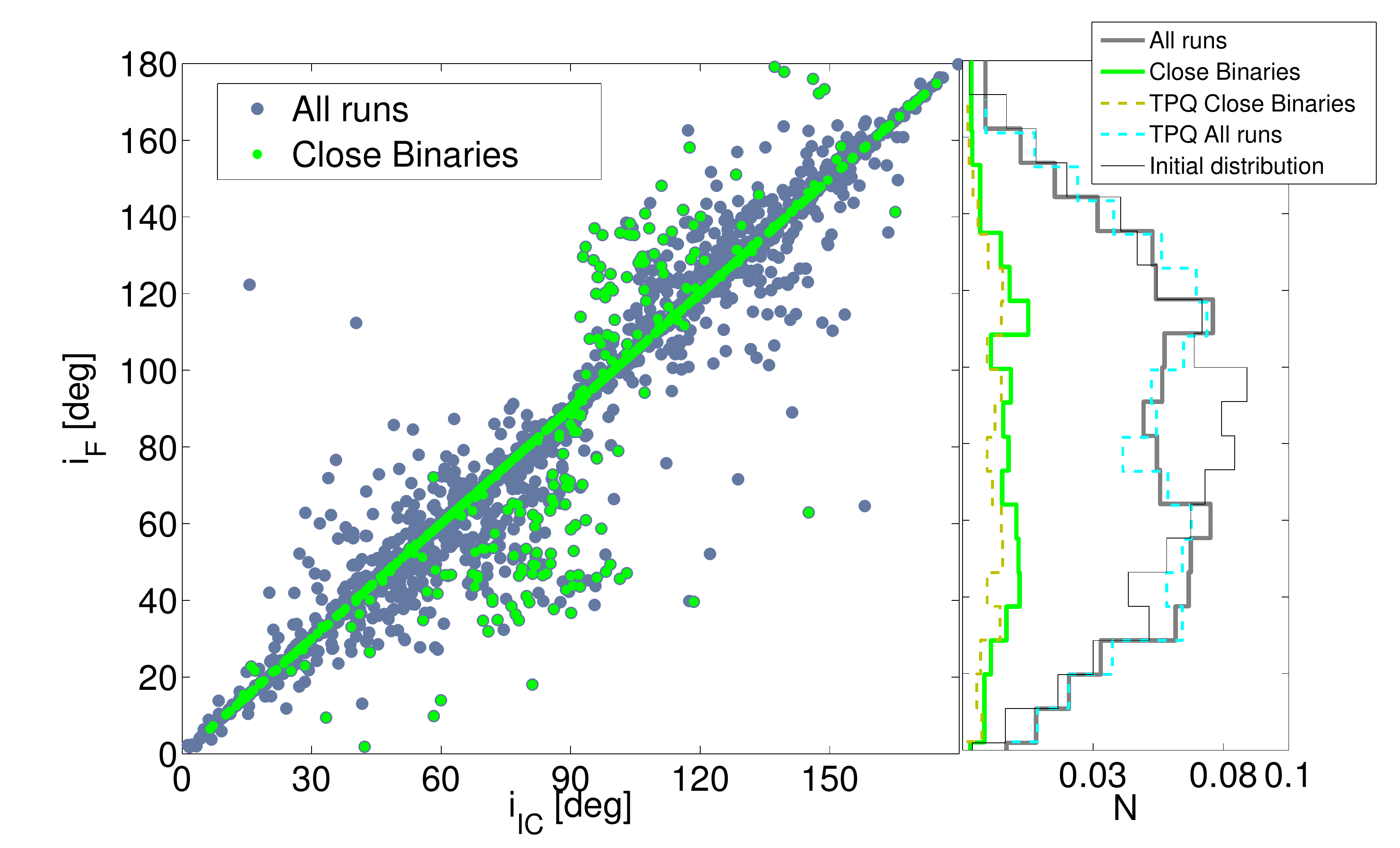}
\caption{  Left panel: the final mutual inclination ($i_F$) as a function of the initial inclination ($i_{IC}$).   The Right panel shows the distribution  of $i_F$ (normalized such that the integral of the ``all runs" distribution  is one) . We consider all of the runs (without those that crossed the Roche--limit), grey,  the close binaries (green)  and the initial distribution (thin black line). We also show in this panel the final inclination distribution of all of the runs and close binaries for the TPQ case (dot-dashed cyan and dark green respectively).
}
   \label{fig:inc} 
\end{figure}

\begin{figure*}[htb]
\begin{center}
\includegraphics[width=\linewidth]{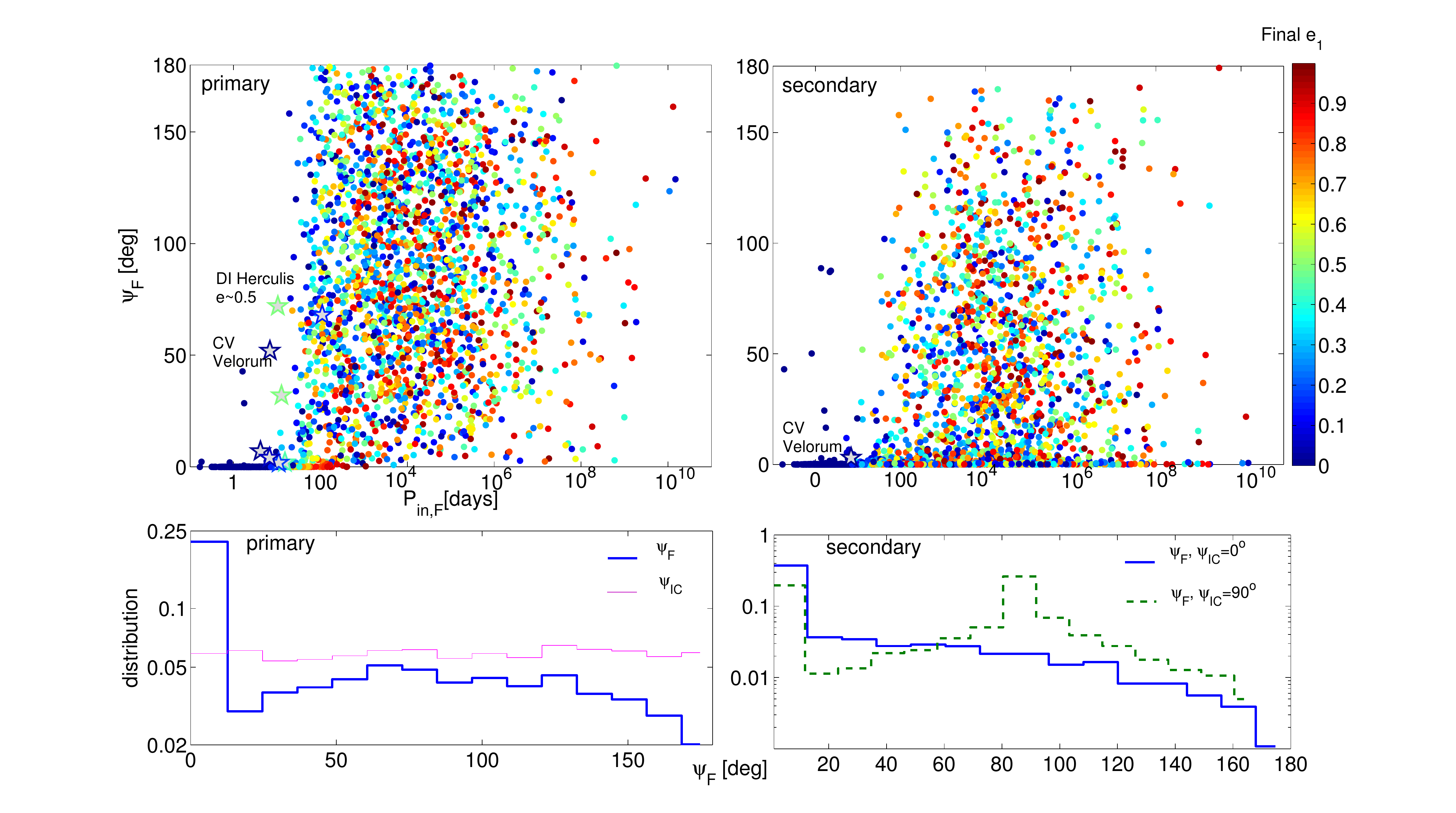}
\caption{Spin orbit distribution of the inner binary. {\bf Top panels:} Final distribution of the spin orbit angle of the primary ({\bf left panel}) and secondary ({\bf right panel}) Vs the final inner orbit periods, the color code is the final eccentricity of the inner binary. We also plot the observations \citep{Albrecht09,Albrecht11,Albrecht13,Albrecht+14,Triaud+13,Harding+13,Zhou+13}. {\bf Bottom panels:} The distribution of the spin orbit angle,  $\psi$, (blue lines) for the inner orbit's primary ({\bf left}) and secondary ({\bf right}) members. The initial distribution of $\psi$ for the primary  was uniform (magenta line in the left panel), while the spin orbit angle of the secondary was set to zero initially.  Note that we have repeated the Monte--Carlo runs, setting initially  the primary spin orbit angle to be aligned just like the secondary and confirmed  that  in that case, the final distribution is identical to the secondary one (omitted here to avoid clatter). Furthermore, to explore the effect of the initial conditions, we have had another Monte--Carlo run setting initially the two spin orbit angles on a perpendicular configuration ($\psi=90^\circ$). For this case we again found that the primary and secondary final spin orbit distribution is identical, and shown in the right bottom panel (green dashed line). Note that both vertical axis in the bottom panels are in logarithmic scale.} \label{fig:spinOrbit}
\end{center}
\end{figure*}

\subsection{Inclination and Spin-Orbit angle}\label{sec:Inc_Spin}

The statistical distribution of mutual orbital inclinations and the spin-orbit angle can help disentangle between different formation scenarios. If the formation involves a chaotic process one may expect an isotropic distribution of mutual inclinations of hierarchal triple. This assumption means that the third body formation is essentially uncorrelated with the inner orbit. Our numerical experiments assumed an isotropic distribution for the initial inclination angle (i.e., uniform in $\cos i$).

 We find, similarly to \citet{Dan}, that the initial inclination distribution is not conserved during the secular evolution. 
We illustrate this  in Figure \ref{fig:inc}, left panel, where we show the final mutual inclination distribution as a function of the initial inclination.  However, unlike  the TPQ results presented in \citet{Dan}, their figure 7, the final inclinations are not confined  to the initial prograde or retrograde configurations, but instead are scattered in the phase space (Figure \ref{fig:inc} left hand panel). This is because the EKL mechanism allows the orbits to flip from $i<90^\circ$ to $i>90^\circ$ and vice versa \citep{Naoz11,Naoz+11sec}. 

The final inclination distribution  is shown in Figure  \ref{fig:inc}, right panel for the close binaries and for all of the runs (for both the EKL and TPQ cases). \citet{Tey+13} showed that  in the absence of dissipation and for initial circular inner orbit, the final distribution of hierarchical triple mutual inclination has in fact three peaks, at $40^\circ,90^\circ$ and $140^\circ$.  The  significance of these peaks  depends on the initial conditions, and low eccentric outer orbits give rise to another peak at $90^\circ$, where the $90^\circ$ is  more dominate, see their figure 14, top panels.     The systems  that reached $90^\circ$ are typically associated with large eccentricity excursions and thus are more likely to undergone tidal evolution. Thus, this peak is absent in the present of dissipation. In other words, the distribution near polar configurations is {\bf slightly more} diluted, which can have large implications for the evolution of compact objects after stellar evolution that requires nearly perpendicular orbits.  

The distribution show in Figure  \ref{fig:inc} yields less prominent peaks at $40^\circ$ and $140^\circ$ as predicted before. This is a result of two main reasons: (1) relaxing the test particle approximation (2) using the EKL approximation. However, we note, that as shown in \citet{Naoz+12bin}  a test particle (such as a hot Jupiter) set initially on a circular orbit results in a final inclination distribution similar to the one predicted by \citet{Dan} .Thus, the distribution with  two main peaks at  $40^\circ$ and $140^\circ$, is recovered in the test particle case for the octupole level of approximation (although with slightly less significance).

Another interesting observable (and perhaps more promising than mutual inclination observations)  is the spin orbit angle (obliquity), which is the angle between the spin axis of the star and the angular momentum of the inner orbit. During the tidal evolution  the obliquity of the closest binary will most likely decay to zero.  The obliquity decays faster than the eccentricity. This process produces inner binaries that are still in the process of shrinking and circularizing with typically low obliquities, as depicted in Figure \ref{fig:P1FP0} bottom panel, where  most  of the simulated binaries below the constant angular momentum run, have small obliquities ($\psi<50^\circ$ for the primary and $\psi<100^\circ$ for the secondary, see below for more details). This behavior is also apparent in the top panel of Figure~\ref{fig:spinOrbit}, where we show the final obliquity distribution. The inner binaries with an intermediate period ($\sim 100$~d) have large eccentricities and low obliquities.

\begin{figure}[!t]
\includegraphics[width=\linewidth]{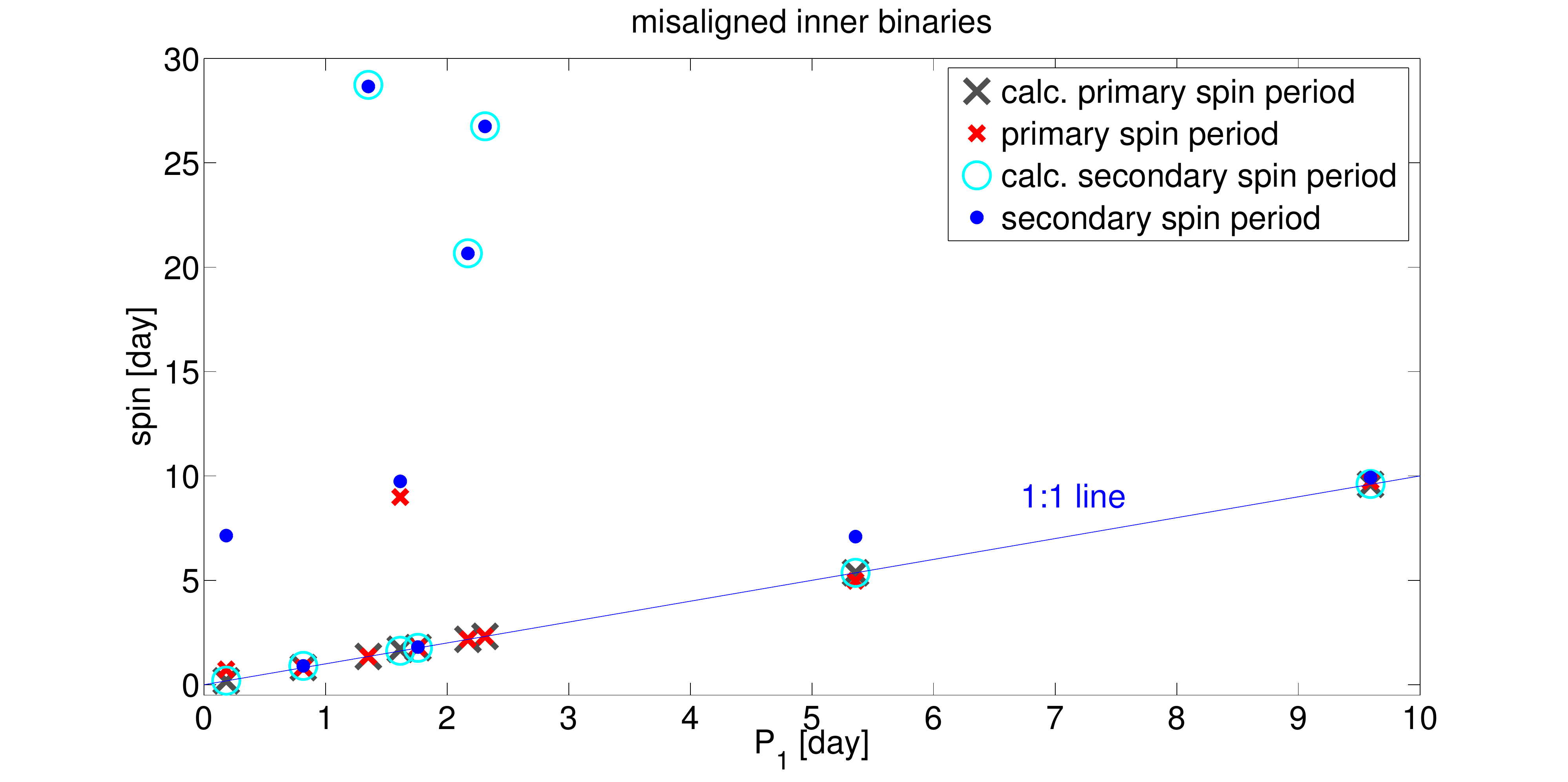}
\caption{  The misalign close inner binaries from Figure \ref{fig:spinOrbit}. We show the simulations results spin periods of the primary (small red crossed) and secondary (small blue dots) as a function of period, for those close ($P_{in}<10~d$) systems that have large obliquity ($>10^\circ$). We also calculate the expected spin period from \citet{Fabrycky+07}, for the primary (large grey crosses) and secondary (open cyan circles).}
   \label{fig:spinsP} 
\end{figure}

 We draw attention to the top two panels of Figure~\ref{fig:spinOrbit}, where there are 9 close ($P_{in}<10$~d) circular  systems with non-negligible  obliquity ($>10$
deg).  All but one of these systems have the primary spin period matching the orbital period, but only 3 of the 9 have secondaries which are similarly synchronized.  The others are spinning considerably slower.  About half of these have experienced chaotic EKL evolution, such that the outcome is a sensitive function of the initial conditions. 
 We show the spin period of these systems for both the primary and secondary in Figure \ref{fig:spinsP}. We also calculate the expected spin period from \citet{Fabrycky+07} :
\begin{equation}\label{eq:spinP}
\Omega=2 \frac{2\pi /P_{in}}{\cos \psi + \sec \psi} \ ,
\end{equation}
which yields  that tilted systems will have smaller spin period \citep[see also][]{Levrard+07}.  We show in Figure \ref{fig:spinsP} that the expected spin period from this calculation agrees well with the results from the Monte-Carlo for the primary and the secondary. 

Very recently, \citet{Albrecht+14} have reported a system, CV
Velorum, with large obliquity ($ 52\pm6^\circ$ for the primary and $ 3\pm7^\circ$ for the secondary).  In Figure \ref{fig:spinOrbit} top panels, this system is located  near the other simulated systems that have very short periods and large obliquities. This may suggest that  CV
Velorum is a result of three-body evolution. Furthermore,  the rotation period of the two stars in this system is similar  to the orbital period of about  $7$~d. This implies that using Equation (\ref{eq:spinP}) one can use the rotation period to constrain the obliquity (or vice versa). For example, the $3^\circ$ value for the secondary obliquity agrees well with having a secondary spin period equal to the orbital period (the $1:1$ line in Figure \ref{fig:spinsP}). However, for the primary we find that a lower obliquity than the mean value (closer to the lower limit, i.e., $46^\circ$) yields an agreement between the calculated spin period and the observed one.

An important parameter is the initial value of the obliquity. One may expect that tight binaries may be well-aligned 
since they originate from the same portion of a molecular cloud. On the other hand, since the formation scenario of even tight binaries involves many chaotic processes, the knowledge of birth obliquity may be unknown. Therefore, we have chosen the following experiments: in the first numerical runs we set initially the primary obliquity from a  {\it uniform} distribution  (uniform in $\psi$)  while the  secondary was chosen to be aligned with the inner orbital angular momentum (i.e., $\psi=0^\circ$). This way we can examine two different initial conditions in the same run. In the second Monte--Carlo test we have set initially the two inner orbit members on a perpendicular configuration, (i.e.,   $\psi=90^\circ$). As a consistency test we have also had a third Monte--Carlo run setting initially both the primary and the secondary on an aligned configuration and confirmed that the final obliquity distributions are identical to the secondary final distribution from the first test.  Note that all the rest of the results shown in this paper are independent on the  initial choice of the obliquity.  For consistency reasons the other orbital parameters discussed and analyzed in the paper belong to the result from the first Monte--Carlo test. 

Different  initial obliquity distributions are distinctive since the final distribution  carries a signature for the initial setup distribution. However, another subtle difference arises in the location of the ``edge"  of the aligned systems, i.e., the smallest period for which most of the systems are aligned.
For example the outcome of the initial {\it perpendicular obliquity} case is that large obliquity systems can extend to very small periods with an approximate limit at $\sim 10$~d, while for the initial {\it zero obliquity} case, final aligned systems can be found for $\lsim 40$~d.

\begin{figure*}[t!]
\includegraphics[width=\linewidth]{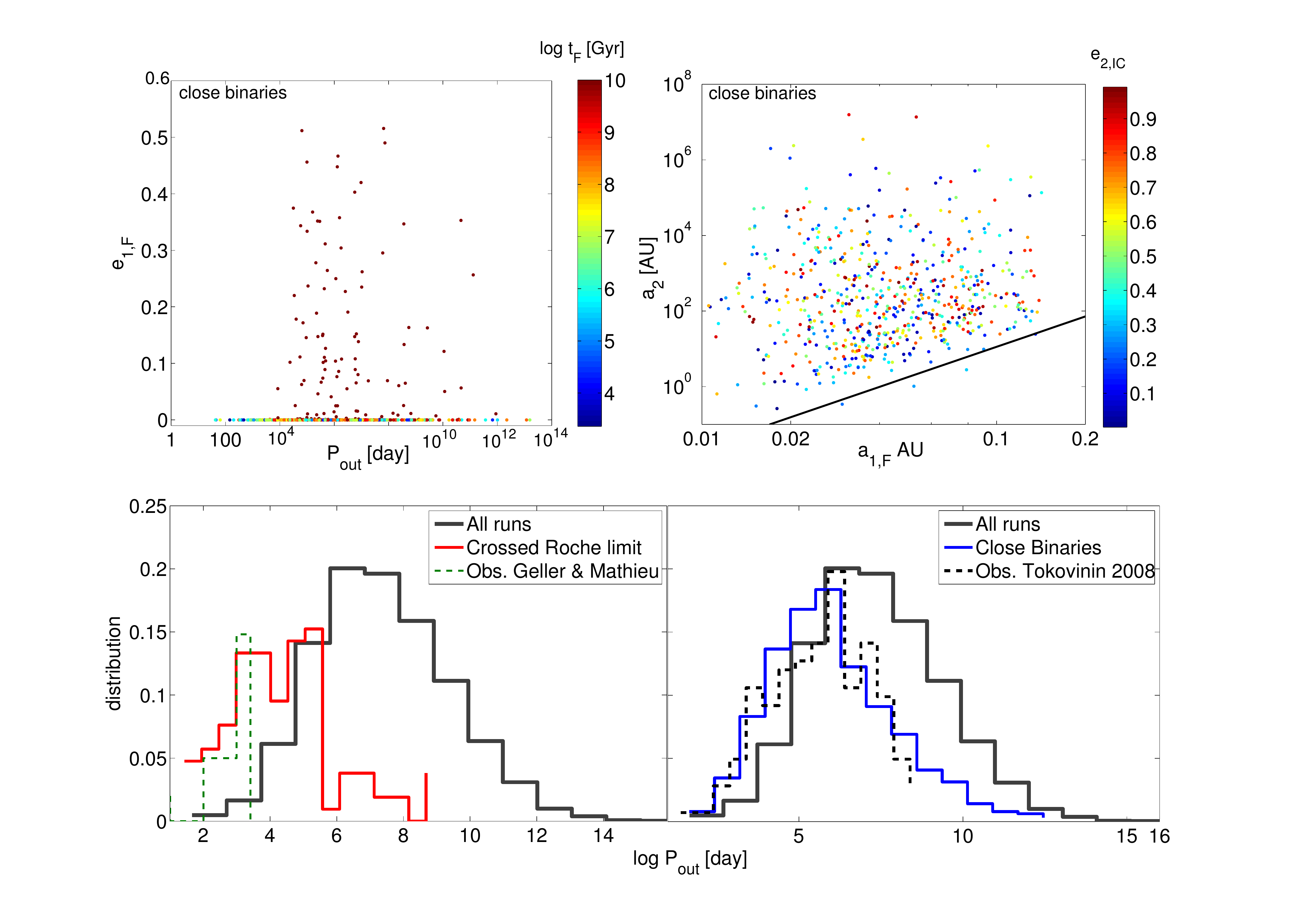}
\caption{ The outer orbit configuration. {\bf Right top panel}: The outer orbit separation ($a_2$, which does not change during the evolution) VS the final inner orbit separation $a_{1,F}$, for the close binaries population. The different colors mark the initial outer orbital eccentricity (which is on average is larger than $0.5$). Note that the outer orbital eccentricity does not change significantly for the population of close binaries (see text). In solid line we mark the  analytical trend for similar mass systems using Eq.~(\ref{eq:SMA_rel}).   Note that about $\sim 50\%$  of the systems (those with $a_2\lsim 100$~AU) started out with $a_1>0.1$~AU  and  have migrated to the left side of the theoretical line.  {\bf Left top panel}: the inner orbit final eccentricity, $e_{1,F}$ as a function of the outer orbit period $P_{out}$, for the close binaries population.  The different colors marks the time the integration stopped. This shows that those systems with large final eccentricity  are still undergoing tidal evolution, and have systematic larger  outer orbit period ($\gsim 10^4$~d)).  {\bf Bottom panels}: the outer orbit period distribution. In both panels we consider all of the runs (blue--grey line). {\bf Right bottom panel}: we show the period of the companion of the close binaries (blue line)  the observed distribution, scaled to match the theory lines, \citep{Tok08}. {\bf Left bottom panel}: we show the period of the merged population (red line) and the observed blue straggles distribution of NGC 188 \citep{GM12}, also scaled to match the theory lines.  
\vspace{1cm}}
   \label{fig:outer} 
\end{figure*}

We found that the final obliquity distribution carries a clear signature of the initial distribution. The final distribution  consists of an aligned population for the tighter binaries, and a distribution which contains an information of the initial setting between $0$ and $40$ days.  This is most apparent for the second Monte--Carlo test (where the $m_1$ and $m_2$ obliquities were initially set to be $90^\circ$). The final distribution has an aligned component,  a broad component of obliquity up to $180^\circ$ , but it also retained a peak at $90^\circ$; see the right bottom panel of Figure  \ref{fig:spinOrbit}. Setting the obliquity initially on a aligned configuration (either for the primary or the secondary) results in a final distribution of an aligned peak with a broad tail of obliquities (see  right bottom panel).

There is a slight difference in setting initially  $m_1$ and $m_2$ on an aligned (perpendicular) configuration compared to setting $m_1$ on a uniform distribution and keeping $m_2$ aligned  (the second run). Apart from having different final obliquity  distributions for the two cases (as depicted in the bottom panels of Figure \ref{fig:spinOrbit}),  the ``edge" of systems on a short period with low obliquities is slightly pushed inward for the case of $m_1$ initially aligned, compared to the uniform $m_1$. This can be  seen by comparing $m_1$ and $m_2$ obliquities for the population with intermediate periods ($\sim 10-100$~d), and non negligible eccentricity, see the two top panels in Figure  \ref{fig:spinOrbit}.
In other words those systems, (at periods of $\sim 10-100$~d with moderate-to high eccentricity and low obliquities), may end up with larger obliquities  when  initially set on an aligned configuration  (the ``blue stripe" of systems in the left top panel, which is absent in the top right panel). This is easily understood if we consider the influence of the  inclination oscillations due to the Kozai mechanism.  During this evolution the inner and outer argument of periapsis    sweeps across $\sim 180^\circ$ which causes large amplitude oscillations on the obliquity as well \citep[see][]{Li+13}. Starting with  zero obliquity, thus can cause larger amplitude oscillations and slightly suppress its damping (because of the larger torque). 
It is interesting to note that DI Herculis \citep{Albrecht09} resides in this ``edge" of intermediate  periods (see top left  panel in Figure  \ref{fig:spinOrbit}).  This suggests DI Herculis misalignment may actually be typical.

\subsection{The outer orbit}\label{sec:outer}
The outer orbit gravitational perturbations can cause  large eccentricity oscillations for the inner orbit, as discussed above. Strong perturbations can result in shrinking of the inner orbit separation or even lead to merger of the inner members (see section \ref{sec:close_inner} for details regarding the inner orbit's properties). The outer orbit configuration  sets  the different outcomes of the inner orbit, and thus a promising observable is  the outer orbit's period distribution. 
 In the bottom right panel of Figure \ref{fig:outer} we show the initial outer orbit period taken from the log-normal distribution of
\citet{Duquennoy+91}. We also show the period distribution of the population of outer orbits that produced close inner binaries (blue line). Interestingly    this distribution is completely  different from the injected initial distribution.  We also show in this Figure the observed outer orbit's distribution  \citep[adopted  from][ public catalog]{Tok08}, which we scaled to guide the eye. Note that the observed distribution is not limited to the close binaries. However, probably  due to observational limitations in compiling the catalog, we suspect that it will be biased toward  companions that are around the close binaries population. Thus, is not surprising that the  Kolmogorov-Smirnov test does not reject the null hypothesis at $5\%$ significance level that the observed outer orbit's distribution and the simulated one are from the same continuous distribution, with $p$ value of $0.1564$.  Therefore, although we have a long tail of wide outer orbits, close binaries $\lsim 16$~d, have preferentially wide outer orbits, with a peak distribution at $\sim 10^6$~day (as shown in Figure \ref{fig:outer} bottom right panel), in agreement with \citet{Tok08} Figure 3.

The close inner binary's final separation represents the final stage of the secular evolution in the presence of tides and GR, where  the outer orbit's separation  does not change.  
When the EKL precession timescale is comparable to tidal (and  GR) precession, further eccentricity excitations from the EKL evolution are suppressed. 
The inner orbit  then settles on the separation that equalized these timescales, or shorter separations.  
The timescale associated with the Newtonian quadrupole  term,  due to the outer body, can be estimated from the canonical equations \citep[e.g.,][]{Naoz+11sec}.
\begin{equation}\label{eq:tquad}
t_{\rm quad}\sim \frac{2\pi a_2^3 (1-e_2^2)^{3/2}\sqrt{m_1+m_2}}{ a_1^{3/2} m_3 \sqrt{G}}  \ ,
\end{equation}
where $G$ is the gravitational constant. The tidal  precession timescale (ignoring the spin term) is estimated as \citep[e.g.][]{1998EKH} 
\begin{equation}\label{eq:tTF}
t_{\rm TF}\sim  \frac{a_1^{13/2} (1-e_1^2)^5 m_1 m_2  }{15 \sqrt{G} \sqrt{m_1+m_2} f_e \Lambda} \ ,
\end{equation}
where $f_e=1 +3 /2   e_1^2 + 1/ 8 e_1^4$ and 
\begin{equation}\label{eq:Lambda}
\Lambda=m_2^2k_{L,1}R_1^5+m_1^2k_{L,2}R_2^5 \ ,
\end{equation}
where $R_j$ and $k_{L,j}$ is the radius and the apsidal motion constant respectively of the $j\in{1,2}$ object in the inner binary.
Equating these two timescales, and solving for $a_1$, we further simplifying this by taking $m_1\sim m_2$  gives the following relation between the two semi major axes:
\begin{equation}\label{eq:SMA_rel}
a_2^3\sim a_1^8\frac{m_3}{m_1} \frac{(1-e_1^2)^5}{60 R_1^5 k_{L,1} f_e (1-e_2)^{3/2}} \ .
\end{equation}
This relation can be used to constrain the other parameters in the problem, for a given inner and outer separations.

\begin{figure}[!t]
\includegraphics[width=9cm,clip=true]{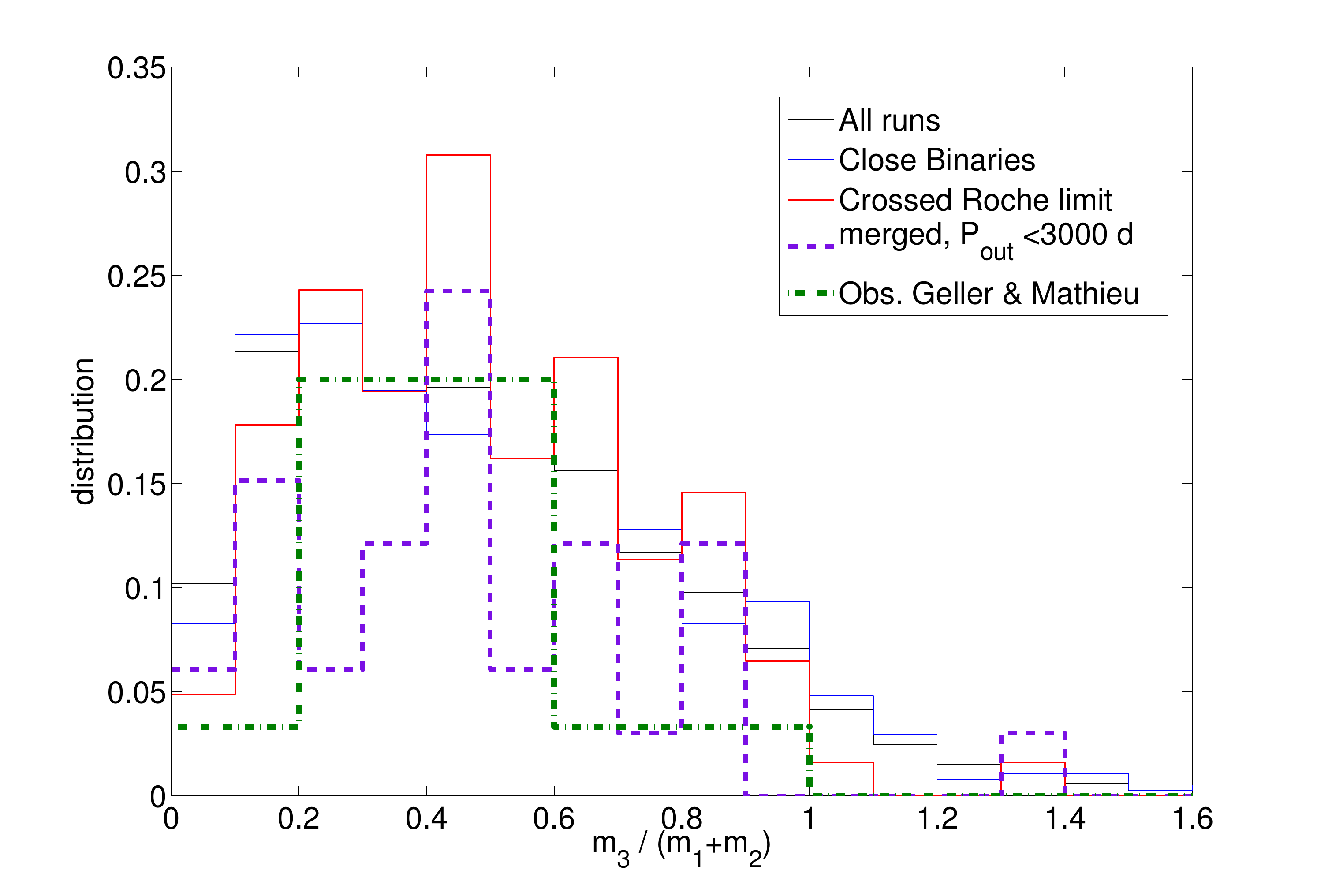}
\caption{ {\bf The distribution of the mass ratio between the outer perturber and the inner binary ($m_3/(m_1+m_2)$). }
 We consider the distribution of all of the runs (blue--grey line) and the close binaries (blue line) and the distribution of the orbits that crossed their Roche limit (red line), which are our designated merged systems. We also plot the observed distribution of the blue-stragglers  for NGC 188 adopted from \citet{GM12} (dashed-dot green line), these of course are binary mass ratio and not triple systems. To compare with observations we also consider a sub-set of systems with $p_{out}<3000$~d (purple dash  line). Note that the observation distribution was rescaled to guide the eye. \vspace{0.5cm}
}
   \label{fig:massRatio} 
\end{figure}

We further approximate Equation  (\ref{eq:SMA_rel}) by taking $m_3\sim m_1$, $e_1\to 0.5$ (which is the mean initial distribution) and $e_2\to 0.5$, since most of the systems that produced close binaries had $e_2\sim 0.5$ (see Figure \ref{fig:outer} top right panel). We get the limit of the relation between the final inner and outer orbit separations (solid line in the top panel of Figure \ref{fig:outer}).  Considering the entire population, $a_2$ seems  uncorrelated  with the final inner orbit separation, $a_{1,F}$, \citep[in agreement with][]{Tok08}. However this limiting line suggest that  different masses and eccentricities   will results in different  relations. In Figure \ref{fig:outer}, about $50\%$ from the systems with $a_2\lsim100$~AU (the relevant systems for this theoretical line) have started to the right side of this line (i.e., with $a_1$ initial larger then the final configuration).

In the left top panel of Figure  \ref{fig:outer}  we show the final inner orbital eccentricity, $e_{1,F}$ vs. the outer orbit period, $P_{out}$, for the close binaries population.  As depicted here, systems with outer orbit period below $\sim 30$~yr have a circular inner orbit. The inner binaries that are still undergoing tidal circularization even after $10$~Gyr of the integration time (i.e., have non negligible final eccentricity) are naturally more likely to have wider outer orbits.  Although wide outer orbits can also cause the inner orbit to shrink and circularize.  

 Another interesting observable is the outer orbit distribution of the inner systems that merged, shown in the left bottom panel of Figure \ref{fig:outer} (red line).  These systems are now binaries thus $P_{out}$ is the ``new" binary period.  Again as for $P_{out}$ of the inner close binaries, the outer orbit population of those inner binaries that merged, is a  distinct subset of the initial period distribution. Not surprisingly, typically close outer orbits will result in a merger of the inner binaries, but as shown in this Figure, a long tail of very wide outer orbit periods (up to $10^9$~d) can still cause the inner binaries to merge.

 These merger products are thus blue stragglers.  \citet{PF09} suggested a similar mechanism for the formation of blue stragglers, however they mainly envisioned a two-step process, in which three-body dynamics plus tidal dissipation created a close binary, and that binary subsequently merged by magnetic winds or had unstable or efficient mass transfer.   Their mechanism explained the contemporaneous observation of a high fraction of companion stars to blue stragglers \citep{MG09}.  However, more recently \citet{GM11} found secondary masses consistent with the typical mass of a white dwarf ($\sim 0.5 M_\odot$), whereas in the \citet{PF09} scenario and ours, the companion masses simply echo the initial conditions of triples -- see figure \ref{fig:massRatio}.  Also, \citet{Gosnell+14} found the UV light of a white dwarf in several systems, suggesting the remnant of a mass donor, rather than the distant companion of the triple dynamics scenario. 

However, we see several reasons why the case is not yet closed in favor of stable mass transfer.  First, the recent simulations of \citet{Geller+13} suggest not enough blue stragglers are made by the standard prescriptions for mass transfer in the best-studied cluster (NGC 188).  Second, five blue stragglers have no companions detected out to 3000 day orbital periods\citep{MG09}; even if there is a more distant companion, this would be too wide for mass transfer to make a blue straggler.  Third, the secondary stars typically have non-zero eccentricities, which a priori would not be expected after the red-giant phase of one of the stars (\citet{1995Verbunt}; although given the uncertain mass-transfer physics, it may be possible \citet{2009Sepinsky}).  Our mechanism can address these three aspects.  

The general principle of collisions in triples was in \cite{Geller+13}'s N-body model (they specifically saw \citet{Leigh+11LS}'s mechanism of collision during unstable resonant encounters), and they also followed stellar evolution models to account for the mass-transfer systems.  They computed fewer blue stragglers than are actually seen in the cluster, however.  We reiterate two caveats \cite{Geller+13} noted: the model lacked primordial triples, and for the dynamically-formed triples it used a formalism that accounted only for the quadrupole interaction (the prescription of \citet{Mardling+01} within NBODY6).  Having primordial triples may increase the yield of blue stragglers, as the outer binary may be perturbed to higher inclination or eccentricity by passing stars, which then triggers a collision by EKL evolution.  And we have shown that the octupole interaction is much more efficient at generating collisions than the quadrupole alone.  So we suggest these extra components may explain the shortfall in modeled blue stragglers. 

  On this second point, we note that our mechanism could explain many of the detected binaries, but it also naturally produces a range of companion periods which could have periods longer than 3000 days.   The cluster environment should unbind the longest-period blue stragglers binaries.  These points were also made by \cite{Geller+13}. 

\begin{figure}[!t]
\includegraphics[width=9cm,clip=true]{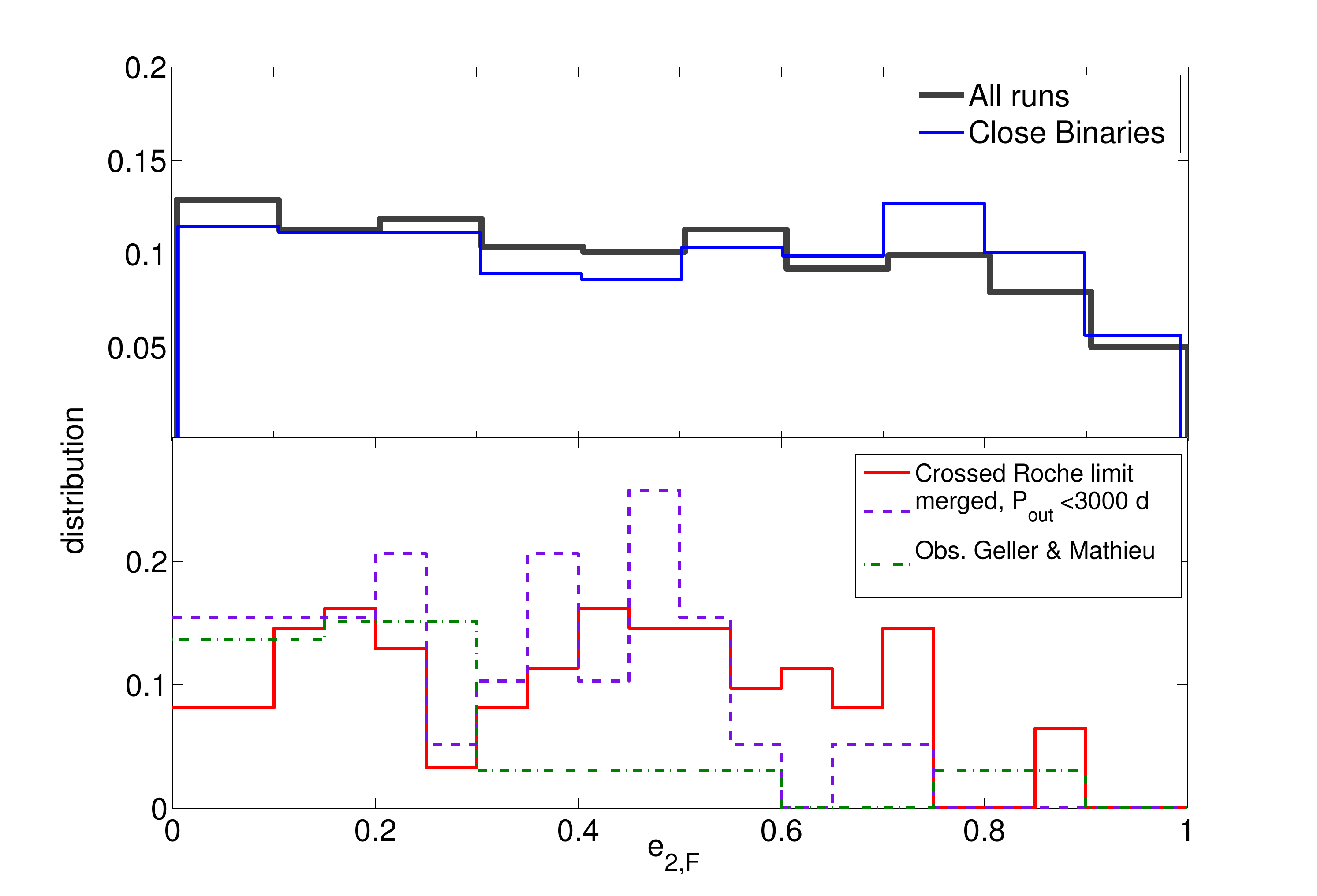}
\caption{ {\bf The distribution of the final outer orbit's eccentricity ($e_{2,F}$).} Note that the final outer orbital eccentricity, almost doesn't change for those systems that ended up either as close binaries or as merged systems.  In the top panel we consider the distribution of all of the runs (blue--grey line) and the closed binaries (blue line). In the bottom panel we consider the distribution of the orbits that crossed their Roche limit (red line), which are our designated merged systems. We also plot the observed distribution of the blue-stragglers  for NGC 188 adopted from \citet{GM12} (dashed-dot green line). To compare with observations we also consider a sub-set of systems with $p_{out}<3000$~d (purple dash line). Note that the observation distribution was rescaled to guide the eye.\vspace{0.5cm}
}
   \label{fig:e2} 
\end{figure}

 Finally, we note that our mechanism naturally predicts an eccentric companion to blue stragglers.  Actually, it appears to produce higher eccentricities than are seen in \citet{MG09}'s observations.   In the bottom panel of Figure \ref{fig:e2} we show the final distribution of $e_2$ of the outer orbit companions for the merged inner systems (red line).  During the EKL evolution the outer orbital eccentricity undergoes small amplitude oscillations \citep{Naoz+11sec}, however, as pointed out by \citet{Tey+13} in the case of comparable mass perturbers, large inner orbital eccentricity are reached when the outer orbital eccentricity almost does not vary. We also find that  the final outer orbital eccentricity almost does not change for those systems that ended up either as close binaries or as merged systems.  Moreover, the merged systems have a distinct $e_{2,F}$ population compared to all of the runs (top panel) that preferentially favor large eccentricities.  We also over-plot the observed eccentricity distribution of the blue stragglers in NGC 188, \citep{GM12}. This distribution is limited to blue-stragglers  binaries with period smaller than $3000$~d. Therefore, to  compare with observations we also consider a sub-set of systems with $P_{out}<3000$~d.

 \section{Discussion}\label{sec:dis}
 
 We have studied the secular evolution of triple stellar systems while considering the octupole level of approximation of the hierarchical three-body problem.  During the system evolution,  the octupole--level of approximation can cause large  eccentricity excitations for the inner orbit. During the large eccentricity excursions,  tidal interactions play an important role. On close pericenter passages, when tides are important the orbital energy is dissipated,  the separation shrinks and the orbit circularizes (see  right panels of Figure \ref{fig:Ex}). The final orbit stabilized on a separation which balances eccentricity excitations from the EKL mechanism and tidal (and/or GR) precession.    On the other hand, if the eccentricity excursion happens on a relative short timescale \citep[but still long so the secular approximation is valid, e.g.,][]{Naoz+11sec}, and tidal (or GR) force cannot influence the dynamics, the binary members may cross each other's Roche limits (see left panels of Figure \ref{fig:Ex}). 
 We considered the systems that crossed their Roche limits as merged systems. 
 
 \begin{itemize}
 \item {\it Comparison with  observations:}\\
 We found that $\sim 21\%$ of all our runs ended up with $p_{in}\lsim 16$~d, and $4\%$ of all the systems crossed the Roche limit (the  merged systems).  We find that the final inner orbit's period distribution agrees with the observed distribution adopted from \citet{Tok08} public catalog  (see Figure   \ref{fig:Pdis}). Furthermore, the inner members configurations that resulted in close binaries (merged systems) represent a distinct sub-set population from the initial injected binary distribution, which constrains the birth properties of close binaries (merged systems).     This also suggests that these subsets may have  only weak dependency on  the  properties of the initial  injected distribution  of all triples.  
 As shown in Figure \ref{fig:Pdis} the close systems had an initial inner period peak of $\sim 10$~d, while the merged systems had an initial period peak of $\sim 1000$~d. Both subsets have a long tail of wide systems that can undergo separation shrinking, ending up either as a close binary or a merged system, where wider inner binaries a slightly more likely to end up in a merged configuration than a close binary.

 An interesting consequence of our results, and specifically the bimodal distribution in Figure \ref{fig:Pdis}, is that only relatively wide inner binaries ($\gsim 16-40$~d) are available for the EKL evolution at the white dwarf  stage (where mass loss will tend to widen the configuration even further).  Furthermore, many of these will not be in a near polar configuration (see Figure \ref{fig:inc}). Many of those close binaries  have already  decoupled from the third star (see Figure \ref{fig:P1FP0}) and are unlikely to undergo large eccentricity excitations at  a later stage. Thus, this distribution should be taken into account for the probability estimations of the  type Ia double degenerate scenario through triple body evolution \citep[e.g.,][]{Tho10,Prodan+13,Katz+12,Dong+14}.

  \citet{Tok08} showed that $P_{out}/P_{in}$ have large values, and concluded the Kozai (TPQ) eccentricity excitations  are suppressed due to general relativistic precessions and that therefore this mechanism cannot produce tight binaries with a wide  outer perturber. Here we claim the exact opposite and supported it by qualitative  comparison  with observations. The EKL mechanism in the presence of tides naturally produces very tight binaries with a companion on a large range of periods (see Figures \ref{fig:P1FP0} and \ref{fig:outer}). When the inner orbit is longer initially, GR does not quench eccentricity excitations and thus tidal dissipation can still take place.  Furthermore, the EKL mechanism, compared to the TPQ case,  extends the valley in the period distribution to larger values with a rather flat period distribution for the close binaries as seen in observations (see Figure \ref{fig:Pdis} and discussion in Section \ref{results}).  
 
  We also find that the outer orbital period distribution is consistent with  observations, both for close inner binary systems (see Figure \ref{fig:outer} right bottom panel, where we compared to   \citet{Tok08} catalog of observed triples) and for the companion of the merged population adopted from \citet{GM12}, see Figure \ref{fig:outer} left bottom panel. The strong agreement with observations for both of those populations emphasize the notion that three body secular interactions may be the main channel for close inner binaries and merged systems like blue stragglers. 
   Future observations can further help  test this idea\footnote{ Note that if this mechanism of the formation of close binaries is the
dominating channel if means that during the inner orbital shrinking
planets test particles (such as planets) my be ejected from the
system. Thus, we would expect a deficit of circumbinary planets in
agreement with observations \citep[e.g.,][]{Armstrong+14}.}.

   \item {\it The Implications of the mechanism on Blue Stragglers formation:}\\ 
    The two main mechanisms that have been proposed in the literature to explain the formation blue stragglers are  coalescence and mass--transfer between two stars \citep{McCrea64}, or collision between the stellar members in a binary either in the field  \citep{Hills+76} or induced by a gravitational perturbations of the third object \citep{PF09}. The latter mechanism is especially promising in explaining blue stragglers binaries. Here we found that about $4\%$ of our runs have crossed each others Roche limit. Both mechanisms may contribute, however here we focus on the secular interactions, and specifically the EKL mechanism.

   The merger between the inner orbit's members typically happens after $~5-100$ quadrupole time scales, Eq.~(\ref{eq:tquad}). In the example shown in Figure \ref{fig:Ex} left panel, the merger happened after $\sim 15$ quadrupole timescales, however, for nearly coplanar systems, the large eccentricity peak can happen on after just a few quadrupole timescales \citep{Li+13}. Regardless  of the exact number of quadrupole timescales before a merge, the merge is not immediate.  
   This implies that the stars typically  will be already on the main sequence phase, furthermore the members will cross the Roche limit during a large eccentricity phase. Therefore, during the evolution we expect an electromagnetic signature that will result from the large velocities (due to the large eccentricities) of the two main sequence stars. 
   
Given these typical numbers, most of the systems that crossed their Roche limit did so in less than few tens of Myr. Therefore, without cluster dynamics included, this mechanism would not make blue stragglers in an open cluster with the age of a few Gyr.  However, primordial triples may be torqued to higher mutual inclination and eccentricity at some point during the life of the cluster, which would then lead to EKL oscillations and a merger, and thus an observable blue straggler.  Measuring  the triple fraction  for open clusters may support this claim that the EKL takes place for newly-formed \emph{or newly-perturbed} triples. However, the estimations of  triples from observations suffer from incompleteness and place a lower limit 
that ranges from $0.5\%$ to $2.3\%$ for different clusters and observations \citep[e.g.][]{Mermilliod+92,Mermilliod+99,Geller+10}.  N-body calculations showed that triples can be dynamically generated on the course  of 7 Gyrs to about $3.8\%$ and maximum of $4.5\%$ \citep{Geller+13}. Furthermore, comparing the type and configurations of the  N-body simulated triples to the observations led  \citet{Geller+13} to conclude that open clusters may form with a significant  population of primordial triples, and they are continuing to form dynamically. Thus, for our purposes, new triples are being formed throughout the life time for the cluster, allowing for the EKL mechanism to take place again for each new triple, which can produce  blue stragglers. Thus the blue stragglers we observe now are recently formed.

 A caveat for these calculations lays in our tidal model. The exact fraction of merged systems depends on the tidal model, and the time scales assumed. \citet{Naoz+12bin} showed that the final fraction of hot Jupiters that merged into their stars depends on the viscous time assumed in the model. Specifically, having shorter viscous time scales by two orders of magnitude compared to the nominal one, resulted in zero merged systems, where two orders of magnitude longer time scales resulted in half as many merged systems. It is reasonable to assume that similar variations will occur here.  

 Recently \citet{Gosnell+14} reported the detection of three young white dwarfs companions to blue stragglers in the NGC 188 star cluster. Interestingly, one of the binaries has a large period ($\sim 3030$~d) and could not be explained as a simple mass transfer or wind accretion  binary. We suggest that this binary may be a result of the dynamical interaction discussed here. The relatively  short age of the white dwarf may suggest that this system formed not too long ago. \citet{Gosnell+14}  suggested that the other two binaries were formed through mass transfer and common envelope  episodes and needed an almost unity mass ratio between the two members. This is rather surprising, since if these two binaries are selected randomly their mass ratio should not be one.  One possibility is that indeed blue stragglers  have a unique mass function which is close to twins. The second possibility is that even the short period detections are an evidence to a dynamical origin.

   \item {\it Implications on the Obliquity Distribution:}\\
 Considering the systems that ended up in a close systems ($\sim 16$~d), we predict a specific distribution for the mutual inclination of the orbit, see Figure \ref{fig:inc}. This distribution is a specific signature for secular three body evolution of the system.  
 Another promising observable is the obliquity of the inner binaries. We found that the final obliquity distribution has a  signature of the initial properties (see Figure \ref{fig:spinOrbit}), which can be used as a tool to study the formation conditions of close binaries in triples. Thus the obliquity  distribution has an aligned peak. Furthermore, we suggested  that observed misaligned binaries such as CV Velorum \citep{Albrecht+14} and DI Herculis \citep{Albrecht09} may have a perturber on a wide orbit, as their current period and obliquities values consistent with the predicted obliquity distribution of our simulated triple systems. 
 
 We have run three Monte-Carlo runs, differ by the initial obliquities of the the inner binary. We found that most of closest binaries have aligned configurations, while the wider ones have a broad obliquity distribution which results from the initial condition. This is most apparent for the Monte-Carlo test which was set initially with a perpendicular configuration.  This case  final distribution, for intermediate--to--long periods  is consistent with a broad distribution with a clear peak at $90^\circ$; see the right bottom panel of Figure  \ref{fig:spinOrbit}. Thus, obtaining a large sample of observed obliquities distributions can help shed  light on the formations property of those systems. 
  \end{itemize}

\section*{Acknowledgments}
We thank Aaron Geller for useful discussions, suggestions and answering many questions. We also thank Tassos Fragos,  Ramesh Narayan, Amaury Triaud, Todd Thompson and Simon Albrecht  for useful discussions. We also thank Andrei Tokovinin for discussing  our arguments about  the comparison between theory and observations.   In addition, we thank our anonymous referee for a thorough reading of the paper and providing many useful comments and suggestions. 
 SN is  supported by NASA through a Einstein Post--doctoral Fellowship awarded by the Chandra X-ray Center, which is operated by the Smithsonian Astrophysical Observatory for NASA under contract PF2-130096.

\bibliographystyle{hapj}

\bibliography{Kozai}

\begin{thebibliography}{85}
\expandafter\ifx\csname natexlab\endcsname\relax\def\natexlab#1{#1}\fi

\bibitem[{{Albrecht} {et~al.}(2009){Albrecht}, {Reffert}, {Snellen}, \&
  {Winn}}]{Albrecht09}
{Albrecht}, S., {Reffert}, S., {Snellen}, I.~A.~G., \& {Winn}, J.~N. 2009,
  \nat, 461, 373, 0909.2861

\bibitem[{{Albrecht} {et~al.}(2013){Albrecht}, {Setiawan}, {Torres},
  {Fabrycky}, \& {Winn}}]{Albrecht13}
{Albrecht}, S., {Setiawan}, J., {Torres}, G., {Fabrycky}, D.~C., \& {Winn},
  J.~N. 2013, \apj, 767, 32, 1211.7065

\bibitem[{{Albrecht} {et~al.}(2011){Albrecht}, {Winn}, {Carter}, {Snellen}, \&
  {de Mooij}}]{Albrecht11}
{Albrecht}, S., {Winn}, J.~N., {Carter}, J.~A., {Snellen}, I.~A.~G., \& {de
  Mooij}, E.~J.~W. 2011, \apj, 726, 68, 1011.0425

\bibitem[{{Albrecht} {et~al.}(2012){Albrecht}, {Winn}, {Fabrycky}, {Torres}, \&
  {Setiawan}}]{Banana}
{Albrecht}, S., {Winn}, J.~N., {Fabrycky}, D.~C., {Torres}, G., \& {Setiawan},
  J. 2012, in IAU Symposium, Vol. 282, IAU Symposium, ed. M.~T. {Richards} \&
  I.~{Hubeny}, 397--398

\bibitem[{{Albrecht} {et~al.}(2014){Albrecht}, {Winn}, {Torres}, {Fabrycky},
  {Setiawan}, {Gillon}, {Jehin}, {Triaud}, {Queloz}, {Snellen}, \&
  {Eggleton}}]{Albrecht+14}
{Albrecht}, S. {et~al.} 2014, ArXiv e-prints, 1403.0583

\bibitem[{{Antognini} {et~al.}(2013){Antognini}, {Shappee}, {Thompson}, \&
  {Amaro-Seoane}}]{Antognini+13}
{Antognini}, J.~M., {Shappee}, B.~J., {Thompson}, T.~A., \& {Amaro-Seoane}, P.
  2013, ArXiv e-prints, 1308.5682

\bibitem[{{Antonini} {et~al.}(2014){Antonini}, {Murray}, \&
  {Mikkola}}]{Antonini+14}
{Antonini}, F., {Murray}, N., \& {Mikkola}, S. 2014, \apj, 781, 45, 1308.3674

\bibitem[{{Armstrong} {et~al.}(2014){Armstrong}, {Osborn}, {Brown}, {Faedi},
  {G{\'o}mez Maqueo Chew}, {Martin}, {Pollacco}, \& {Udry}}]{Armstrong+14}
{Armstrong}, D.~J., {Osborn}, H., {Brown}, D., {Faedi}, F., {G{\'o}mez Maqueo
  Chew}, Y., {Martin}, D., {Pollacco}, D., \& {Udry}, S. 2014, ArXiv e-prints,
  1404.5617

\bibitem[{{Blaes} {et~al.}(2002){Blaes}, {Lee}, \& {Socrates}}]{Bla+02}
{Blaes}, O., {Lee}, M.~H., \& {Socrates}, A. 2002, \apj, 578, 775,
  arXiv:astro-ph/0203370

\bibitem[{{Bode} \& {Wegg}(2014)}]{Bode+14}
{Bode}, J.~N., \& {Wegg}, C. 2014, \mnras, 438, 573

\bibitem[{{Conroy} {et~al.}(2013){Conroy}, {Prsa}, {Stassun}, {Orosz},
  {Fabrycky}, \& {Welsh}}]{Conroy+13}
{Conroy}, K.~E., {Prsa}, A., {Stassun}, K.~G., {Orosz}, J.~A., {Fabrycky},
  D.~C., \& {Welsh}, W.~F. 2013, ArXiv e-prints, 1306.0512

\bibitem[{{Dong} {et~al.}(2014){Dong}, {Katz}, {Kushnir}, \&
  {Prieto}}]{Dong+14}
{Dong}, S., {Katz}, B., {Kushnir}, D., \& {Prieto}, J.~L. 2014, ArXiv e-prints,
  1401.3347

\bibitem[{{Duquennoy} \& {Mayor}(1991)}]{Duquennoy+91}
{Duquennoy}, A., \& {Mayor}, M. 1991, \aap, 248, 485

\bibitem[{{Eggleton}(1983)}]{Eggleton83}
{Eggleton}, P.~P. 1983, \apj, 268, 368

\bibitem[{{Eggleton} {et~al.}(1998){Eggleton}, {Kiseleva}, \& {Hut}}]{1998EKH}
{Eggleton}, P.~P., {Kiseleva}, L.~G., \& {Hut}, P. 1998, \apj, 499, 853,
  arXiv:astro-ph/9801246

\bibitem[{{Eggleton} \& {Kiseleva-Eggleton}(2001)}]{Egg+01}
{Eggleton}, P.~P., \& {Kiseleva-Eggleton}, L. 2001, \apj, 562, 1012,
  arXiv:astro-ph/0104126

\bibitem[{{Eggleton} \& {Kisseleva-Eggleton}(2006)}]{Egg+06}
{Eggleton}, P.~P., \& {Kisseleva-Eggleton}, L. 2006, \apss, 304, 75

\bibitem[{{Eggleton} {et~al.}(2007){Eggleton}, {Kisseleva-Eggleton}, \&
  {Dearborn}}]{Eggleton+07}
{Eggleton}, P.~P., {Kisseleva-Eggleton}, L., \& {Dearborn}, X. 2007, in IAU
  Symposium, Vol. 240, IAU Symposium, ed. {W.~I.~Hartkopf, E.~F.~Guinan, \&
  P.~Harmanec}, 347--355

\bibitem[{{Fabrycky} \& {Tremaine}(2007)}]{Dan}
{Fabrycky}, D., \& {Tremaine}, S. 2007, \apj, 669, 1298, 0705.4285

\bibitem[{{Fabrycky} {et~al.}(2007){Fabrycky}, {Johnson}, \&
  {Goodman}}]{Fabrycky+07}
{Fabrycky}, D.~C., {Johnson}, E.~T., \& {Goodman}, J. 2007, \apj, 665, 754,
  astro-ph/0703418

\bibitem[{{Ford} {et~al.}(2000){Ford}, {Kozinsky}, \& {Rasio}}]{Ford00}
{Ford}, E.~B., {Kozinsky}, B., \& {Rasio}, F.~A. 2000, \apj, 535, 385

\bibitem[{{Geller} {et~al.}(2010){Geller}, {Hurley}, \& {Mathieu}}]{Geller+10}
{Geller}, A.~M., {Hurley}, J.~R., \& {Mathieu}, R.~D. 2010, in IAU Symposium,
  Vol. 266, IAU Symposium, ed. R.~{de Grijs} \& J.~R.~D. {L{\'e}pine},
  258--263, 0911.4382

\bibitem[{{Geller} {et~al.}(2011){Geller}, {Hurley}, \& {Mathieu}}]{aaron+11}
{Geller}, A.~M., {Hurley}, J.~R., \& {Mathieu}, R.~D. 2011, in Bulletin of the
  American Astronomical Society, Vol.~43, American Astronomical Society Meeting
  Abstracts \#217, 327.02--+

\bibitem[{{Geller} {et~al.}(2013){Geller}, {Hurley}, \& {Mathieu}}]{Geller+13}
{Geller}, A.~M., {Hurley}, J.~R., \& {Mathieu}, R.~D. 2013, \aj, 145, 8,
  1210.1575

\bibitem[{{Geller} \& {Mathieu}(2011)}]{GM11}
{Geller}, A.~M., \& {Mathieu}, R.~D. 2011, \nat, 478, 356, 1110.3793

\bibitem[{{Geller} \& {Mathieu}(2012)}]{GM12}
------. 2012, \aj, 144, 54, 1111.3950

\bibitem[{{Goldreich} \& {Soter}(1966)}]{GS66}
{Goldreich}, P., \& {Soter}, S. 1966, \icarus, 5, 375

\bibitem[{{Gosnell} {et~al.}(2014){Gosnell}, {Mathieu}, {Geller}, {Sills},
  {Leigh}, \& {Knigge}}]{Gosnell+14}
{Gosnell}, N.~M., {Mathieu}, R.~D., {Geller}, A.~M., {Sills}, A., {Leigh}, N.,
  \& {Knigge}, C. 2014, ArXiv e-prints, 1401.7670

\bibitem[{{Griffin}(2012)}]{Griffin12}
{Griffin}, R.~F. 2012, Journal of Astrophysics and Astronomy, 33, 29

\bibitem[{{Hamers} {et~al.}(2013){Hamers}, {Pols}, {Claeys}, \&
  {Nelemans}}]{Hamers+13}
{Hamers}, A.~S., {Pols}, O.~R., {Claeys}, J.~S.~W., \& {Nelemans}, G. 2013,
  \mnras, 430, 2262, 1301.1469

\bibitem[{{Hansen}(2010)}]{Hansen10}
{Hansen}, B.~M.~S. 2010, \apj, 723, 285, 1009.3027

\bibitem[{{Harding} {et~al.}(2013){Harding}, {Hallinan}, {Konopacky},
  {Kratter}, {Boyle}, {Butler}, \& {Golden}}]{Harding+13}
{Harding}, L.~K., {Hallinan}, G., {Konopacky}, Q.~M., {Kratter}, K.~M.,
  {Boyle}, R.~P., {Butler}, R.~F., \& {Golden}, A. 2013, \aap, 554, A113,
  1304.5290

\bibitem[{{Harrington}(1968)}]{Har68}
{Harrington}, R.~S. 1968, \aj, 73, 190

\bibitem[{{Harrington}(1969)}]{Har69}
------. 1969, Celestial Mechanics, 1, 200

\bibitem[{{Hills} \& {Day}(1976)}]{Hills+76}
{Hills}, J.~G., \& {Day}, C.~A. 1976, \aplett, 17, 87

\bibitem[{{Holman} {et~al.}(1997){Holman}, {Touma}, \& {Tremaine}}]{Hol+97}
{Holman}, M., {Touma}, J., \& {Tremaine}, S. 1997, \nat, 386, 254

\bibitem[{{Hut}(1980)}]{Hut}
{Hut}, P. 1980, \aap, 92, 167

\bibitem[{{Ivanova} {et~al.}(2010){Ivanova}, {Chaichenets}, {Fregeau},
  {Heinke}, {Lombardi}, \& {Woods}}]{Iva+10}
{Ivanova}, N., {Chaichenets}, S., {Fregeau}, J., {Heinke}, C.~O., {Lombardi},
  J.~C., \& {Woods}, T.~E. 2010, \apj, 717, 948, 1001.1767

\bibitem[{{Katz} \& {Dong}(2012)}]{Katz+12}
{Katz}, B., \& {Dong}, S. 2012, ArXiv e-prints, 1211.4584

\bibitem[{{Katz} {et~al.}(2011){Katz}, {Dong}, \& {Malhotra}}]{Boaz2}
{Katz}, B., {Dong}, S., \& {Malhotra}, R. 2011, ArXiv e-prints, 1106.3340

\bibitem[{{Kiseleva} {et~al.}(1998){Kiseleva}, {Eggleton}, \&
  {Mikkola}}]{1998KEM}
{Kiseleva}, L.~G., {Eggleton}, P.~P., \& {Mikkola}, S. 1998, MNRAS, 300, 292

\bibitem[{{Kozai}(1962)}]{Kozai}
{Kozai}, Y. 1962, \aj, 67, 591

\bibitem[{{Kushnir} {et~al.}(2013){Kushnir}, {Katz}, {Dong}, {Livne}, \&
  {Fern{\'a}ndez}}]{Kushnir+13}
{Kushnir}, D., {Katz}, B., {Dong}, S., {Livne}, E., \& {Fern{\'a}ndez}, R.
  2013, \apjl, 778, L37, 1303.1180

\bibitem[{{Leigh} \& {Sills}(2011)}]{Leigh+11LS}
{Leigh}, N., \& {Sills}, A. 2011, \mnras, 410, 2370, 1009.0461

\bibitem[{{Leigh} \& {Geller}(2013)}]{LG13}
{Leigh}, N.~W.~C., \& {Geller}, A.~M. 2013, \mnras, 432, 2474, 1304.2775

\bibitem[{{Levrard} {et~al.}(2007){Levrard}, {Correia}, {Chabrier}, {Baraffe},
  {Selsis}, \& {Laskar}}]{Levrard+07}
{Levrard}, B., {Correia}, A.~C.~M., {Chabrier}, G., {Baraffe}, I., {Selsis},
  F., \& {Laskar}, J. 2007, \aap, 462, L5, astro-ph/0612044

\bibitem[{{Li} {et~al.}(2014){Li}, {Naoz}, {Holman}, \& {Loeb}}]{Li+14}
{Li}, G., {Naoz}, S., {Holman}, M., \& {Loeb}, A. 2014, ArXiv e-prints,
  1405.0494

\bibitem[{{Li} {et~al.}(2013){Li}, {Naoz}, {Kocsis}, \& {Loeb}}]{Li+13}
{Li}, G., {Naoz}, S., {Kocsis}, B., \& {Loeb}, A. 2013, ArXiv e-prints,
  1310.6044

\bibitem[{{Lidov}(1962)}]{Lidov}
{Lidov}, M.~L. 1962, planss, 9, 719

\bibitem[{{Lithwick} \& {Naoz}(2011)}]{LN}
{Lithwick}, Y., \& {Naoz}, S. 2011, \apj, 742, 94, 1106.3329

\bibitem[{{Mardling} \& {Aarseth}(2001)}]{Mardling+01}
{Mardling}, R.~A., \& {Aarseth}, S.~J. 2001, \mnras, 321, 398

\bibitem[{{Mathieu} \& {Geller}(2009)}]{MG09}
{Mathieu}, R.~D., \& {Geller}, A.~M. 2009, \nat, 462, 1032

\bibitem[{{Mazeh} \& {Shaham}(1979)}]{Mazeh+79}
{Mazeh}, T., \& {Shaham}, J. 1979, AA, 77, 145

\bibitem[{{McCrea}(1964)}]{McCrea64}
{McCrea}, W.~H. 1964, \mnras, 128, 147

\bibitem[{{Mermilliod} \& {Mayor}(1999)}]{Mermilliod+99}
{Mermilliod}, J.-C., \& {Mayor}, M. 1999, \aap, 352, 479,
  arXiv:astro-ph/9911405

\bibitem[{{Mermilliod} {et~al.}(1992){Mermilliod}, {Rosvick}, {Duquennoy}, \&
  {Mayor}}]{Mermilliod+92}
{Mermilliod}, J.-C., {Rosvick}, J.~M., {Duquennoy}, A., \& {Mayor}, M. 1992,
  \aap, 265, 513

\bibitem[{{Michaely} \& {Perets}(2014)}]{Michaely+14}
{Michaely}, E., \& {Perets}, H.~B. 2014, ArXiv e-prints, 1406.3035

\bibitem[{{Miller} \& {Hamilton}(2002)}]{MH02}
{Miller}, M.~C., \& {Hamilton}, D.~P. 2002, \apj, 576, 894,
  arXiv:astro-ph/0202298

\bibitem[{{Naoz} {et~al.}(2011){Naoz}, {Farr}, {Lithwick}, {Rasio}, \&
  {Teyssandier}}]{Naoz11}
{Naoz}, S., {Farr}, W.~M., {Lithwick}, Y., {Rasio}, F.~A., \& {Teyssandier}, J.
  2011, \nat, 473, 187, 1011.2501

\bibitem[{{Naoz} {et~al.}(2013{\natexlab{a}}){Naoz}, {Farr}, {Lithwick},
  {Rasio}, \& {Teyssandier}}]{Naoz+11sec}
------. 2013{\natexlab{a}}, \mnras, 431, 2155, 1107.2414

\bibitem[{{Naoz} {et~al.}(2012){Naoz}, {Farr}, \& {Rasio}}]{Naoz+12bin}
{Naoz}, S., {Farr}, W.~M., \& {Rasio}, F.~A. 2012, \apjl, 754, L36, 1206.3529

\bibitem[{{Naoz} {et~al.}(2013{\natexlab{b}}){Naoz}, {Kocsis}, {Loeb}, \&
  {Yunes}}]{Naoz+12GR}
{Naoz}, S., {Kocsis}, B., {Loeb}, A., \& {Yunes}, N. 2013{\natexlab{b}}, \apj,
  773, 187, 1206.4316

\bibitem[{{Pejcha} {et~al.}(2013){Pejcha}, {Antognini}, {Shappee}, \&
  {Thompson}}]{Pejcha+13}
{Pejcha}, O., {Antognini}, J.~M., {Shappee}, B.~J., \& {Thompson}, T.~A. 2013,
  \mnras, 435, 943, 1304.3152

\bibitem[{{Perets}(2014)}]{Perets14}
{Perets}, H.~B. 2014, ArXiv e-prints, 1406.3490

\bibitem[{{Perets} \& {Fabrycky}(2009)}]{PF09}
{Perets}, H.~B., \& {Fabrycky}, D.~C. 2009, \apj, 697, 1048, 0901.4328

\bibitem[{{Perets} \& {Kratter}(2012)}]{PK12}
{Perets}, H.~B., \& {Kratter}, K.~M. 2012, ArXiv e-prints, 1203.2914

\bibitem[{{Pribulla} \& {Rucinski}(2006)}]{Pri+06}
{Pribulla}, T., \& {Rucinski}, S.~M. 2006, \aj, 131, 2986,
  arXiv:astro-ph/0601610

\bibitem[{{Prodan} {et~al.}(2013){Prodan}, {Murray}, \& {Thompson}}]{Prodan+13}
{Prodan}, S., {Murray}, N., \& {Thompson}, T.~A. 2013, ArXiv e-prints,
  1305.2191

\bibitem[{{Raghavan} {et~al.}(2010){Raghavan}, {McAlister}, {Henry}, {Latham},
  {Marcy}, {Mason}, {Gies}, {White}, \& {ten Brummelaar}}]{Raghavan+10}
{Raghavan}, D. {et~al.} 2010, \apjs, 190, 1, 1007.0414

\bibitem[{{Rappaport} {et~al.}(2013){Rappaport}, {Deck}, {Levine}, {Borkovits},
  {Carter}, {El Mellah}, {Sanchis-Ojeda}, \& {Kalomeni}}]{Rappaport+13}
{Rappaport}, S., {Deck}, K., {Levine}, A., {Borkovits}, T., {Carter}, J., {El
  Mellah}, I., {Sanchis-Ojeda}, R., \& {Kalomeni}, B. 2013, \apj, 768, 33,
  1302.0563

\bibitem[{{Sepinsky} {et~al.}(2009){Sepinsky}, {Willems}, {Kalogera}, \&
  {Rasio}}]{2009Sepinsky}
{Sepinsky}, J.~F., {Willems}, B., {Kalogera}, V., \& {Rasio}, F.~A. 2009, \apj,
  702, 1387, 0903.0621

\bibitem[{{Shappee} \& {Thompson}(2013)}]{Shappee+13}
{Shappee}, B.~J., \& {Thompson}, T.~A. 2013, \apj, 766, 64, 1204.1053

\bibitem[{{Teyssandier} {et~al.}(2013){Teyssandier}, {Naoz}, {Lizarraga}, \&
  {Rasio}}]{Tey+13}
{Teyssandier}, J., {Naoz}, S., {Lizarraga}, I., \& {Rasio}, F. 2013, ArXiv
  e-prints, 1310.5048

\bibitem[{{Thompson}(2011)}]{Tho10}
{Thompson}, T.~A. 2011, \apj, 741, 82, 1011.4322

\bibitem[{{Tokovinin}(2008)}]{Tok08}
{Tokovinin}, A. 2008, \mnras, 389, 925, 0806.3263

\bibitem[{{Tokovinin}(2014{\natexlab{a}})}]{Tokovinin14b}
------. 2014{\natexlab{a}}, \aj, 147, 86, 1401.6825

\bibitem[{{Tokovinin}(2014{\natexlab{b}})}]{Tokovinin14a}
------. 2014{\natexlab{b}}, \aj, 147, 87, 1401.6827

\bibitem[{{Tokovinin} {et~al.}(2006){Tokovinin}, {Thomas}, {Sterzik}, \&
  {Udry}}]{Tokovinin+06}
{Tokovinin}, A., {Thomas}, S., {Sterzik}, M., \& {Udry}, S. 2006, \aap, 450,
  681, astro-ph/0601518

\bibitem[{{Tokovinin}(1997)}]{T97}
{Tokovinin}, A.~A. 1997, Astronomy Letters, 23, 727

\bibitem[{{Tokovinin} \& {Smekhov}(2002)}]{TS02}
{Tokovinin}, A.~A., \& {Smekhov}, M.~G. 2002, \aap, 382, 118

\bibitem[{{Triaud} {et~al.}(2013){Triaud}, {Hebb}, {Anderson}, {Cargile},
  {Collier Cameron}, {Doyle}, {Faedi}, {Gillon}, {Gomez Maqueo Chew},
  {Hellier}, {Jehin}, {Maxted}, {Naef}, {Pepe}, {Pollacco}, {Queloz},
  {S{\'e}gransan}, {Smalley}, {Stassun}, {Udry}, \& {West}}]{Triaud+13}
{Triaud}, A.~H.~M.~J. {et~al.} 2013, \aap, 549, A18, 1208.4940

\bibitem[{{Veras} \& {Tout}(2012)}]{Veras+12}
{Veras}, D., \& {Tout}, C.~A. 2012, \mnras, 422, 1648, 1202.3139

\bibitem[{{Verbunt} \& {Phinney}(1995)}]{1995Verbunt}
{Verbunt}, F., \& {Phinney}, E.~S. 1995, \aap, 296, 709

\bibitem[{{Wen}(2003)}]{Wen}
{Wen}, L. 2003, \apj, 598, 419, arXiv:astro-ph/0211492

\bibitem[{{Zhou} \& {Huang}(2013)}]{Zhou+13}
{Zhou}, G., \& {Huang}, C.~X. 2013, ArXiv e-prints, 1307.2249

\end{thebibliography}

\end{document}